\documentclass{aa}
\usepackage{txfonts}
\usepackage{natbib}
\usepackage{graphicx}
\usepackage{amssymb}
\usepackage{mathrsfs}
\usepackage[dvips]{rotating}
\bibpunct{(}{)}{;}{a}{}{,}

\def\uv{\mathrm{uv}}

\def\bol{\mathrm{bol}}

\def\AU{\hbox{AU}}

\def\eff{\mathrm{eff}}

\def\A{\r{A} }

\def\OI{[O\,{\sc i}] }
\def\SII{\hbox{S\,{\sc ii}}}

\def\Ha{H$\alpha$ }
\def\kms{$\mathrm{km}~\mathrm{s}^{-1}$}
\def\kep{\mathrm{kep}}
\def\proj{\mathrm{proj}}
\begin{document}
%\thesaurus{02.01.2,08.03.4,08.06.2,08.16.5,13.09.6}
\title{[O\,{\sc i}] 6300\A emission in Herbig Ae/Be systems: \\
signature of Keplerian rotation\thanks{Based on observations collected
at the European Southern Observatory, La Silla, Chile (program numbers
54.D-0363 and 68.C-0348).}
} 
\titlerunning{\OI emission in HAEBE disks.}
\authorrunning{Acke, van~den~Ancker \& Dullemond}
\author{B.~Acke\inst{1} \and M.~E.~van~den~Ancker\inst{2}\fnmsep\thanks{Visiting Astronomer, Kitt
Peak National Observatory, National Optical Astronomy Observatory, which is
operated by the Association of Universities for Research in Astronomy, Inc.
(AURA) under cooperative agreement with the National Science Foundation.} \and
  C.~P.~Dullemond\inst{3}}
\institute{Instituut voor Sterrenkunde, KULeuven, Celestijnenlaan 200B, 
3001 Leuven, Belgium\\ 
\email{Bram.Acke@ster.kuleuven.ac.be}
\and
European Southern Observatory, Karl-Schwarzschild Strasse 2, D-85748
Garching bei M\"unchen, Germany
\and
Max-Planck-Institut f\"ur Astrophysik, Karl-Schwarzschildstrasse 1,
Postfach 1317, D-85748 Garching bei M\"unchen, Germany}
\date{DRAFT, \today}

\abstract{ 
   We present high spectral-resolution optical spectra of 49 Herbig Ae/Be
   stars in a search for the \OI 6300\A  line. The vast
   majority of the stars in our sample show narrow (FWHM $<$ 100 km~s$^{-1}$)
   emission lines, centered on the stellar radial velocity.  In only
   three sources is the feature much broader ($\sim$ 400 km~s$^{-1}$),
   and strongly blueshifted ($-$200 km~s$^{-1}$) compared to the stellar
   radial velocity.  Some stars in our sample show double-peaked lines
   profiles, with peak-to-peak separations of $\sim$ 10 km~s$^{-1}$.
   The presence and strength of the \OI line emission appears
   to be correlated with the far-infrared energy distribution of
   each source: stars with a strong excess at 60~$\mu$m have in general
   stronger [O\,{\sc i}] emission than stars with weaker 60~$\mu$m excesses.
   We interpret these narrow \OI 6300\A line profiles as
   arising in the surface layers of the protoplanetary disks surrounding
   Herbig Ae/Be stars.  A simple model for \OI 6300\A line
   emission due to the photodissociation of OH molecules shows that our
   results are in quantitative agreement with that expected from the
   emission of a flared disk if the fractional OH abundance is
   $\sim$5 $\times$ 10$^{-7}$.
}

\maketitle

%---------------------------------------------------------------------

\section{Introduction}

Herbig Ae/Be (HAEBE) stars are intermediate-mass pre-main-sequence
objects. The spectral energy distribution (SED) of these sources is
characterized by 
the presence of an infrared excess, due to thermal emission of
circumstellar dust. For HAEBE stars of spectral type F, A and late-B,
the evidence for a disk-like geometry of this circumstellar dust
\citep[e.g.][]{mannings97, testi, pietu, fuente, natta04} is
generally accepted. The spatial distribution of the circumstellar
matter around early-B-type stars is less clear
\citep[e.g.][]{natta00}. Early-B stars 
have dissipation time scales for the circumstellar spherical envelope
of the order of their pre-main-sequence life time. 
Nevertheless observations seem to indicate that at
least some of these sources have disks \citep{vink02}.
Both disk-like and spherical geometries might be present around these
sources.

\citet[][henceforth M01]{meeus01} have classified the 14 isolated
HAEBE stars in their sample based on the
shape of the mid-IR (20--100~$\mu$m) SED. \textit{Group~I} contains the
sources which have a rising mid-IR flux excess (\textit{double-peaked}
SED). \textit{Group~II} sources display more modest IR excesses. M01
suggest phenomenologically that the difference in SED shape reflects a
different disk geometry: group~I sources have flared disks, while
group~II objects have geometrically flat disks.

\citet{dullemond02}, \citet{dullemond02b} and
\citet{dullemond04} have modeled circumstellar
disks with a self-consistent model based on 2D-radiative transfer
coupled to the equations of vertical hydrostatics. The model is
composed of a disk with an inner hole ($\sim$ 0.5 \AU), a puffed-up
inner rim and an outer part. The inner rim, which is located at the dust
sublimation temperature, is puffed-up due to the direct head-on stellar
radiation and the consequent increase of the local gas temperature.
The models show that the outer part of the disk can either be flared
\citep[as in the models of ][]{chiang}, or lie completely in
the shadow of 
the puffed-up inner rim. The SED of a flared disk displays a strong
mid-IR excess, comparable to that of the M01 group~I
sources. Self-shadowed disk models have SEDs with a steeper decline
towards longer wavelengths, like the SEDs of group~II sources. The
\citet{dullemond02} models hence connect the M01 classification to the
disk geometry: group~I sources have flared disks, group~II objects
self-shadowed disks. 

\citet[][henceforth AV04]{ackeiso} have investigated all available
near-infrared \textit{Infrared Space Observatory}
\citep[ISO,][]{kessler} spectra of HAEBE 
stars. In their paper, the emission bands at 3.3, 6.2, ``7.7'', 8.6
and 11.3~$\mu$m were studied qualitatively. These emission
features are attributed to \textit{polycyclic aromatic hydrocarbons}
\citep[PAHs,][]{leger,allamandola}. The latter, carbonaceous
molecules, are thought to be 
excited by ultraviolet (UV) photons, and radiate in the near-infrared in
bands linked to the vibrational modes of the CC and CH bonds. 
AV04 have found a correlation between the strength of the PAH
features and the shape of the SED:
group~I sources display significant PAH emission in their spectra. 
Group~II members on the other hand show much
weaker PAH bands, indicating the lack of excited PAH molecules in
these systems. Following the remark of M01, AV04 suggest this is due
to the geometry of the circumstellar disk. In self-shadowed
geometries, the stellar UV 
flux needed to excite the PAH molecules cannot reach the disk surface,
while in flared disks this surface is directly exposed to the radiation
field of the central star. Theoretical modeling of the PAH emission in flared
circumstellar disks \citep{habart} is in agreement with the observations.

Forbidden-line emission from different elements \citep[C {\sc
i}, N {\sc ii}, O {\sc i}, O {\sc ii}, S {\sc ii}, Ca {\sc ii}, Cr
{\sc ii}, Fe {\sc ii}, Ni {\sc ii}; see e.g.,][]{hamann94} in the
optical part of the spectrum has been
observed in a considerable amount of HAEBEs. In the present paper we
focus on the \OI features at 6300 and 6363\A. The circumstellar region from
which these lines emanate has been the subject of a long-standing
debate in the literature \citep[e.g.,][]{finkenzeller85,bohm94,hirth,
hamann94,bohm97,corcoran97,corcoran98,hernandez04}. There is agreement
that the blueshifted high-velocity (a few 100 \kms) wing observed in a
few sources is formed in an outflow whose redshifted part is obscured
by the circumstellar disk. The origin of low-velocity \textit{symmetric} \OI
emission profile is less clear. \citet{kwan} have provided a
qualitative explanation for the observed \OI profiles in T Tauri
stars. Their model consists of a low- and a high-velocity component,
the first due to a slow disk wind, the latter emanating from a
collimated jet. Previously published suggestions for the
forbidden-line emitting region in HAEBE stars are based on this model
and  include a low-velocity disk wind \citep[][]{hirth, corcoran97}
and a spherically 
symmetric stellar wind \citep[\textit{without} an obscuring
  circumstellar disk,][]{bohm94}. In the latter explanation, the
authors suggest that the \OI line is formed in the outermost parts of
the stellar wind. 

With the present paper, we intend to contribute to this discussion,
and try to explain the observed low-velocity component in a broader
framework. In 
\textsection\ref{secdataset}, the composition of the sample is presented.
First, we investigate the observational data and perform a
quantitative analysis. We search for correlations between parameters
describing the forbidden-line emission and the SED. We have also
checked the possible connection between 
the presence and strength of PAH features and the forbidden-line
emission (\textsection\textsection\ref{secanalysis},
\ref{secinterpretation}). Second, we propose a simple model 
for the forbidden-line emission region, and compare the model results
to the observations (\textsection\ref{secmodeling}).

%---------------------------------------------------------------------

\section{The data set \label{secdataset}}

\subsection{The sample \label{secsample}}

We have observed 49 Herbig Ae/Be stars in the wavelength
region around the forbidden oxygen line at 6300\A. Since we intend to
compare the results of the \textit{ISO} study of HAEBE stars by AV04
to the present [O\,{\sc i}]-emission-line analysis, we have compiled
our \OI sample in order to have as large as possible an overlap with
their sample. Additional HAEBE stars from the catalogues of 
\citet[][Table 1 and 2 in their article]{the} and \citet{malfait} were
observed when possible to enlarge the sample. 

Optical spectra of the sample stars were obtained with four different
instruments on five different telescopes: the Coud\'e
Echelle Spectrometer (CES) on the Coud\'e Auxiliary Telescope (CAT)
and on the ESO 3.6m telescope, the Fiber-fed Extended Range Optical
Spectrograph (FEROS) on the ESO 1.5m telescope, the CCD Echelle
Spectrograph on the Mayall 4m 
telescope at Kitt Peak National Observatory (KPNO) and the
Utrecht Echelle Spectrograph (UES) on the William Herschel Telescope
(WHT). The first three telescopes are located at 
the ESO site La Silla (Chile), KPNO is in Arizona (USA) and the
WHT is situated on La Palma (Canary Islands, Spain). The FEROS and
WHT measurements were made by Gwendolyn Meeus. The spectra were
obtained at different periods during the past 10 years. In
Table~\ref{sample}, we have summarized the sample stars and spectra
included in this analysis. For 10 sources, two or more spectra were
obtained.

In order to extract the spectra from the raw data, a standard
echelle-data reduction was applied. This includes background
subtraction, cosmic-hit removal, flatfielding and
wavelength calibration. The spectra were normalized to unity by
fitting a spline function 
through continuum points and dividing by it. The reduction and normalization was
performed using the ESO program MIDAS. 

The spectral resolution $\lambda / \Delta\lambda$ at 6300\r{A}
is instrument-dependent. The resolution for the KPNO data is $\sim$30,000, for
the FEROS and WHT spectra $\sim$45,000, and $\sim$65,000 for the CAT data.
The highest spectral resolution is reached in the ESO 3.6m 
spectra with $\lambda / \Delta\lambda \approx 125,000$. The FEROS and WHT data cover a large
fraction of the optical wavelength range (3700--8860\A and
5220--9110\A respectively), which allows us to accurately determine the radial
velocity of the central star ($v_{rad}$) based on many
photospheric absorption lines 
throughout the spectrum. The KPNO data cover the wavelength range
between 5630 and 6640\r{A}. We have used the Fe {\sc ii} absorption line at
5780.128\A to estimate the stellar radial velocity in these spectra. The CES
spectra (CAT and ESO 3.6m) are limited to one order around the \OI line at 6300\r{A}. No
stellar radial velocities could be determined from these spectra.
For a few sources, Ca {\sc ii} K 3934\r{A} spectra are at our disposal. When no
photospheric-line spectra were available to us, we estimated the radial velocity
based on these circumstellar lines.
An estimate of $v_{rad}$ was retrieved from the literature in case we
could not determine it from any of our spectra. When no literature
estimate exists, we have computed an average $v_{rad}$ of stars in 
the area on the sky (radius 2\arcmin) around the source.
The radial velocities are included in Table~\ref{samplevsOI} (see later).

The spectra were velocity-rebinned and, after applying a heliocentric
correction, centered around the stellar radial velocity.
In this way we can compare the measured velocities, independently of
the intrinsic radial velocity of the entire system.

\begin{table*}[!th]
\caption{ The sample of HAEBEs used in this analysis, based on
  \citet{the} and \citet{malfait}. For each object, the available optical
spectra are indicated. Furthermore, the SIMBAD values for right
ascension (RA) and declination (Dec) are given, as well as the date,
start time and integration time T of the observation. 
\label{sample}}
\begin{center}
\begin{tabular}{ccccccc}
 \multicolumn{7}{c}{\bf Sample stars}\\ \hline \hline
 \multicolumn{1}{c}{Object} &
 \multicolumn{1}{c}{Spectrum} &
 \multicolumn{1}{c}{RA (2000)} &  
 \multicolumn{1}{c}{Dec (2000)} &
 \multicolumn{1}{c}{Date} &
 \multicolumn{1}{c}{Start} &
 \multicolumn{1}{c}{T} \\ 
     &     & $h\hspace{0.3cm} m\hspace{0.3cm} s $ & \degr \hspace{0.2cm} \arcmin \hspace{0.2cm} \arcsec & \textit{dd/mm/yy} & \textit{h\ \ m} &  [m] \\
\hline
   V376 Cas       &  KPNO       & 00 11 26.1 & $+$58 50 04   &     25/06/02 &  10:22  & 20 \\  
   VX Cas	  &  KPNO       & 00 31 30.7 & $+$61 58 51   &     24/06/02 &  10:06  & 25 \\  
   AB Aur         &  FEROS      & 04 55 45.8 & $+$30 33 04   &     17/01/99 &  22:37  & 21 \\  
   HD 31648       &  WHT        & 04 58 46.3 & $+$29 50 37   &     24/12/96 &  00:00  & 20 \\  
   HD 34282       &  ESO 3.6m   & 05 16 00.5 & $-$09 48 35   &     01/04/02 &  00:49  & 50 \\  
   HD 34700       &  ESO 3.6m   & 05 19 41.4 & $+$05 38 43   &     01/04/02 &  00:05  & 40 \\  
   HD 35929       &  CAT        & 05 27 42.8 & $-$08 19 38   &     13/01/94 &  05:27  & 30 \\  
   HD 36112       &  WHT        & 05 30 27.5 & $+$25 19 57   &     24/12/96 &  02:18  & 25 \\  
   HD 244604      &  WHT        & 05 31 57.3 & $+$11 17 41   &     24/12/96 &  02:48  & 20 \\  
   HD 245185      &  WHT        & 05 35 09.6 & $+$10 01 52   &     24/12/96 &  03:42  & 12 \\  
   V586 Ori       &  CAT        & 05 36 59.3 & $-$06 09 16   &     13/01/94 &  03:10  & 60 \\  
   BF Ori         &  CAT        & 05 37 13.3 & $-$06 35 01   &     13/01/94 &  04:22  & 60 \\  
   Z CMa          &  ESO 3.6m   & 07 03 43.2 & $-$11 33 06   &     01/04/02 &  01:57  & 30 \\  
   	          &  WHT	&            &               &     24/12/96 &  05:36  & 20 \\  
   HD 95881       &  ESO 3.6m   & 11 01 57.6 & $-$71 30 48   &     01/04/02 &  23:46  & 17 \\  
   HD 97048       &  ESO 3.6m   & 11 08 03.3 & $-$77 39 18   &     01/04/02 &  22:56  & 20 \\  
                  &  FEROS      &            &               &     28/01/99 &  04:28  & 55 \\  
   HD 98922       &  CAT        & 11 22 31.7 & $-$53 22 12   &     13/01/94 &  08:40  & 15 \\  
   HD 100453      &  ESO 3.6m   & 11 33 05.6 & $-$54 19 29   &     01/04/02 &  01:44  & 10 \\  
                  &  FEROS      &            &               &     26/01/99 &  04:55  & 20 \\  
   HD 100546      &  CAT        & 11 33 25.4 & $-$70 11 41   &     13/01/94 &  07:55  & 10 \\  
                  &  ESO 3.6m   &            &               &     01/04/02 &  02:32  & 10 \\  
                  &  ESO 3.6m   &            &               &     31/03/02 &  23:27  & 10 \\  
                  &  FEROS      &            &               &     28/01/99 &  05:27  & 30 \\  
   HD 101412      &  CAT        & 11 39 44.5 & $-$60 10 28   &     13/01/94 &  09:06  & 20 \\  
   HD 104237      &  ESO 3.6m   & 12 00 05.1 & $-$78 11 35   &     31/03/02 &  23:35  &  5 \\  
                  &  FEROS      &            &               &     27/01/99 &  03:38  & 20 \\  
   HD 135344      &  ESO 3.6m   & 15 15 48.4 & $-$37 09 16   &     01/04/02 &  02:40  & 23 \\  
                  &  FEROS      &            &               &     27/01/99 &  04:02  & 45 \\  
   HD 139614      &  ESO 3.6m   & 15 40 46.4 & $-$42 29 54   &     01/04/02 &  03:06  & 17 \\  
                  &  FEROS      &            &               &     27/01/99 &  04:49  & 30 \\  
   HD 141569      &  ESO 3.6m   & 15 49 57.8 & $-$03 55 16   &     01/04/02 &  04:55  &  6 \\  
                  &  KPNO       &            &               &     23/06/02 &  03:57  & 10 \\  
   HD 142666      &  ESO 3.6m   & 15 56 40.0 & $-$22 01 40   &     01/04/02 &  03:52  & 29 \\  
   HD 142527      &  ESO 3.6m   & 15 56 41.9 & $-$42 19 23   &     01/04/02 &  03:30  & 20 \\  
   HD 144432      &  ESO 3.6m   & 16 06 58.0 & $-$27 43 10   &     01/04/02 &  04:22  & 23 \\  
   HR 5999	  &  ESO 3.6m   & 16 08 34.3 & $-$39 06 18   &     01/04/02 &  04:46  &  6 \\  
   HD 150193      &  ESO 3.6m   & 16 40 17.9 & $-$23 53 45   &     01/04/02 &  05:04  & 30 \\  
   AK Sco         &  ESO 3.6m   & 16 54 44.9 & $-$36 53 19   &     01/04/02 &  05:34  & 29 \\  
CD$-$42$^\circ$11721 & ESO 3.6m & 16 59 06.8 & $-$42 42 08   &     01/04/02 &  06:05  & 60 \\  
   HD 163296      &  ESO 3.6m   & 17 56 21.3 & $-$21 57 22   &     01/04/02 &  07:08  &  5 \\  
   HD 169142      &  ESO 3.6m   & 18 24 29.8 & $-$29 46 49   &     01/04/02 &  07:15  & 17 \\  
   MWC 297	  &  ESO 3.6m   & 18 27 39.6 & $-$03 49 52   &     01/04/02 &  08:43  & 60 \\  
          	  &  KPNO       &            &               &     25/06/02 &  05:40  & 20 \\  
   VV Ser	  &  KPNO       & 18 28 49.0 & $+$00 08 39   &     23/06/02 &  04:49  & 30 \\  
   R CrA          &  CAT        & 19 01 53.7 & $-$36 57 08   &     19/08/94 &  12:43  & 45 \\  
   T Cra	  &  ESO 3.6m   & 19 01 58.8 & $-$36 57 49   &     01/04/02 &  07:39  & 40 \\  
   HD 179218      &  ESO 3.6m   & 19 11 11.6 & $+$15 47 16   &     01/04/02 &  09:47  &  8 \\  
                  &  KPNO       &            &               &     24/06/02 &  04:24  & 10 \\  
   WW Vul         &  KPNO       & 19 25 58.8 & $+$21 12 31   &     23/06/02 &  06:31  & 30 \\
   HD 190073      &  KPNO       & 20 03 02.5 & $+$05 44 17   &     24/06/02 &  05:01  & 10 \\
BD$+$40$^\circ$4124 &  KPNO     & 20 20 28.3 & $+$41 21 52   &     23/06/02 &  08:11  & 20 \\
LkH$\alpha$ 224   &  KPNO       & 20 20 29.3 & $+$41 21 28   &     24/06/02 &  05:40  & 30 \\
LkH$\alpha$ 225   &  KPNO       & 20 20 30.5 & $+$41 21 27   &     23/06/02 &  09:21  & 30 \\
 \hline
\end{tabular}
\end{center}
\end{table*}

\begin{table*}[!th]
\begin{center}
\begin{tabular}{ccccccc}
 \multicolumn{7}{c}{\bf Sample stars (continued)}\\ \hline \hline
 \multicolumn{1}{c}{Object} &
 \multicolumn{1}{c}{Spectrum} &
 \multicolumn{1}{c}{RA (2000)} &  
 \multicolumn{1}{c}{Dec (2000)} &
 \multicolumn{1}{c}{Date} &
 \multicolumn{1}{c}{Start} &
 \multicolumn{1}{c}{T} \\ 
     &     & $h\ \ m\ \ s\ $ & $^{\circ}\ \ \ '\ \ \ "$ & \textit{dd/mm/yy} & \textit{h\ \ m} &  [m]  \\
\hline
   PV Cep	  &  KPNO       & 20 45 54.3 & $+$67 57 39   &     24/06/02 &  07:18  & 30 \\
   HD 200775      &  KPNO       & 21 01 36.9 & $+$68 09 48   &     23/06/02 &  09:56  & 15 \\
   V645 Cyg       &  KPNO       & 21 39 58.0 & $+$50 14 24   &     25/06/02 &  07:12  & 30 \\
BD$+$46$^\circ$3471 &  KPNO     & 21 52 34.1 & $+$47 13 45   &     25/06/02 &  08:52  & 20 \\
   SV Cep	  &  KPNO       & 22 21 33.3 & $+$73 40 18   &     23/06/02 &  10:19  & 20 \\
   IL Cep	  &  KPNO       & 22 53 15.6 & $+$62 08 45   &     25/06/02 &  09:58  & 20 \\
   MWC 1080       &  KPNO       & 23 17 25.6 & $+$60 50 43   &     24/06/02 &  08:59  & 30 \\
 \hline
\end{tabular}
\end{center}
\end{table*}

In this analysis, we focus on the \OI emission line at
6300\r{A}. For the sample stars with FEROS, KPNO and WHT spectra, the
extended spectral range allows us to include the \OI line at
6363\A as well. Furthermore, these spectra contain the \Ha line at
6550\r{A}.

\subsection{Measurement of the \OI emission lines \label{measoi}}

Before analysing the \OI 6300\r{A} line, we have removed the telluric
absorption lines from the spectrum by cutting out these wavelength regions
and replacing them by a spline approximation based on all the
other data points in the spectrum. The airglow feature (\OI emission from the
Earth's atmosphere) was suppressed in the same way.

In ten of our sample sources, the 6300\A region is rich in
photospheric absorption lines. This hampers the detection of possible
superimposed \OI emission. Of the other 39 stars, 29 objects display a
pure-emission shape (i.e. no underlying absorption features). The ten
remaining sources have no detected \OI line.

We have measured the equivalent width
(EW) in \A and centroid position\footnote{$ \mathrm{centroid~position}
  = \frac{\int v I_{cs}(v) dv}{\int I_{cs}(v) dv}$ with $I_{cs}$ the
  continuum-subtracted 
  intensity of the profile and $v$ the velocity \hbox{parameter}.} in
\kms. Following tradition, the EW is \textit{negative} when the
feature is in \textit{emission}. The EW in
absorption-rich 6300\A spectra is determined in a fixed interval
($\pm$50~\kms) around the stellar radial velocity. We remind the
reader that the centroid position is
given with respect to the reference frame of the central star: a
positive (negative) centroid position indicates that the feature is
redshifted (blueshifted) compared to the stellar radial velocity. For
the detected emission lines, we have also 
determined the full width at half maximum (FWHM) by fitting a Gaussian
function to the feature. This estimate of the FWHM is corrupted
by the instrumental profile of the spectrograph.
Since the telluric lines in the spectra are intrinsicly very narrow,
the FWHM of these absorption features is a good estimate for the
instrumental width. The measured FWHM of the \OI lines is corrected for
this instrumental broadening. \textit{All} detected
features in this sample are spectrally resolved (i.e. have a corrected
FWHM several times larger than the instrumental width).
Finally, we have defined an asymmetry
parameter $\mathcal{A}$ based on the centroid position and the extreme
blue ($\mathcal{B}$) and red ($\mathcal{R}$) ends of the emission
profile: $\mathcal{A}=(\mathrm{centroid} - \mathcal{B})/(\mathcal{R}
- \mathrm{centroid})$. When $\mathcal{A}$ is larger (smaller) than 1, the
centroid position lies closer to the red (blue) end of the emission line,
hence indicating a stronger \textit{red (blue)} wing compared to the center
$\mathcal{C}$ of the
line\footnote{$\mathcal{C}=(\mathcal{R}+\mathcal{B})/2$}. Note that the
line profile can be completely blueshifted compared to the stellar radial
velocity, while having $\mathcal{A} = 1$. The only \OI line parameter
that is sensitive to (errors on) the stellar radial velocity is the
centroid position.

Errors on the \OI parameter values have been estimated based on the
noise in the spectrum. In the figures that are presented in this
paper, we have chosen to show a representative error bar instead of
plotting the individual error bars on each measurement. This was done
for the sake of clarity. The plotted error bar, which indicates the errors on
a mean entry in the figure, is refered to in the captions
of the figures as \textit{the typical error bar}.

When no \OI feature was detected, we have computed an upper limit for the
EW from the noise on the data ($\sigma$). We assume that the strongest
feature that remains undetected is a rectangular
emission line with a height of 5 times the noise on the data and a
width of 1\A ($= 47$ \kms). The equivalent width of this hypothetical
feature thus is $\mathrm{EW}_{max} = 5\sigma~ \A$. The latter
value was used as the upper limit for the undetected line.

For the 10 sources for which we have more than one optical spectrum at our
disposal, we have measured the \OI 6300\A feature in all spectra
separately. Afterwards the parameter values 
were compared. No significant differences were noted, except for
\object{Z CMa} (for which the $|\mathrm{EW}|$ increased with $\sim$40\% in 6
years) and \object{MWC 297} (decrease of $|\mathrm{EW}|$ with $\sim$25\% in 3
months). This suggests that for most stars in our  
sample the \OI emission is constant in time, although variations
--either in the continuum flux or in the line emission itself-- do
occur in some objects. The final data consist of one measurement for
each source, which is a weighted average when two or more spectra are
available. The centroid positions of the features in the
different spectra of the same source agree well within the error bar,
except again for \object{MWC 297}, where the difference in centroid
position adds up to 20 \kms. The forbidden-line emission region of
this source apparently rotates \textit{as a whole} around the central
star. 

Thanks to the large wavelength coverage of the FEROS, KPNO and WHT
spectra, also the wavelength region around the forbidden
\OI 6363\A emission line was observed for 30 objects. 
When present in these spectra, we have determined the EW
and FWHM of the \OI 6363\A emission line. Otherwise, an upper limit
for the EW was determined. The EW of the \Ha
line was measured as well. When no \Ha spectrum was available to us,
we have included literature values.

\subsection{The SED parameters}

To characterize the spectral energy distribution (SED) of the
central star, several quantities were determined, based on
UV-to-millimeter photometry from the literature. For each source, a
\citet{kurucz} model with effective temperature $T_\eff$ and surface
gravity $\log g$ corresponding to the
spectral type of the source was fitted to the de-reddened photometry. From this
model, parameters like total
luminosity $L$ and UV luminosity $L_\uv$ (2--13.6 eV) can be
determined. By subtracting the Kurucz model 
from the infrared photometry, the flux excesses at 2.2
(i.e. $K$ band), 60, 850 and 1300 micron can be determined. Other
parameters used in this analysis are the 
the observed bolometric luminosity $L_\bol$ (not corrected for
extinction) and the distance $d$ to the source. 

To convert the measured EW of the \OI lines to \OI
luminosities, we have computed the theoretical 6300\A continuum flux from
the Kurucz model. The relationship between the \OI luminosity
$L$([O\,{\sc i}]) and the EW is 
\begin{equation}
 L(\rm{[O\,{\mathsc i}]}) = 4 \pi d^2~~ 2.5 \times 10^2~~ |\mathrm{EW}| \times
\mathcal{F}(6300\A) 
\end{equation}
where $L$([O\,{\sc i}]) is in $[L_\odot]$, $d$ in $[\mathrm{pc}]$, EW in [\r{A}]
and the continuum flux $\mathcal{F}$ in [W m$^{-2}$ $\mu$m$^{-1}$]. In a similar way,
\Ha luminosities were computed from the \Ha EWs.

\begin{figure}
\rotatebox{0}{\resizebox{3.5in}{!}{\includegraphics{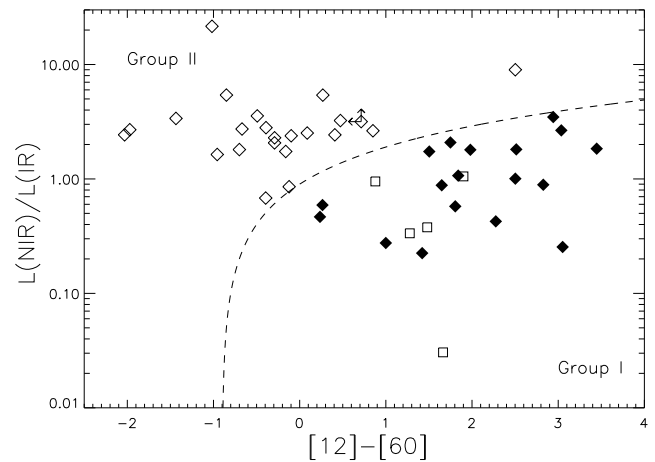}}} 
\caption{ Classification diagram, based on \citet{vanboekel}.
  Objects in the upper left corner are classified as group
  II sources (open diamonds), sources on the lower right side are
  either group~I (filled diamonds) or group~III (squares) members. The
  dashed line represents the empirical separation between group~I and
  II: $L(NIR)/L(IR) = ([12]-[60])+0.9$ (AV04). The quantities
  $L(NIR)/L(IR)$ and $IRAS~[12]-[60]$ are listed in Table~\ref{roytable}.}
\label{plotroyoi.ps}
\end{figure}

The sample sources were classified into the M01 groups, based on the
shape of their IR excess. In accordance with \citet{vanboekel}, we
characterize the SED using the following quantities: the ratio of
$L(NIR)$ (the integrated luminosity derived from broad-band
$J$, $H$, $K$, $L$ and $M$ photometry) and $L(IR)$ (the corresponding
quantity derived from $IRAS$ 12, 25 and 60 micron points), and the
non-color-corrected $IRAS~[12]-[60]$ color. These quantities naturally
separate sources with a strong mid-IR excess (group~I) from
more modest mid-IR emitters (group~II). In Fig.~\ref{plotroyoi.ps},
the classification is visualized in a diagram. The dashed line
represents $L(NIR)/L(IR) = ([12]-[60])+0.9$, which empirically
provides the best separation between both groups. This classification
method has also been used in \citet{dullemond03} and AV04 and has
been described more thoroughly in \citet{ackesubmm}. The quantities
$L(NIR)/L(IR)$ and $[12]-[60]$ are listed in Table~\ref{roytable}.

Five sources in this sample display the amorphous 10 micron silicate
feature in absorption (AV04). Considering the presence of an outflow and
CO bandhead emission, strong veiling and a UV excess, these objects
are thought to have disks whose luminosity is dominated by viscous
dissipation of energy due to accretion, and are deeply embedded
sources. This idea is supported by their very red IR spectral energy
distribution. They are likely fundamentally different from the other
sample stars. We therefore classify them in a 
different group: \textit{group~III}.

BD$+$40$^\circ$4124, \hbox{R CrA} and \hbox{Lk\Ha 224} have not been
classified based on their location in the diagram in
Fig.~\ref{plotroyoi.ps}. Confusion with background sources in the $IRAS$
photometry made it impossible to derive the desired quantities
$L(NIR)/L(IR)$ and $[12]-[60]$. BD$+$40$^\circ$4124 has been classified
as a group~I source, because its SED resembles that of HD
200775. \hbox{R CrA} and \hbox{Lk\Ha 224} are both UX Orionis
stars. According to \citet{dullemond03}, these sources are group~II
sources. Hence we have classified them as members of the latter group.

\begin{table}
\caption{ The non-color-corrected $IRAS~[12]-[60]$ color and the
  near-to-mid IR luminosity ratio $L(NIR)/L(IR)$ of the sample
  sources.
\label{roytable}}
\begin{center}
\begin{tabular}{lcc}
Object & $IRAS~[12]-[60]$ & $L(NIR)/L(IR)$ \\
\hline
\textbf{Group I}     &  & \\
V376 Cas     &      1.42  &    0.22  \\
AB Aur       &      1.50  &    1.73 \\
HD 34282     &      3.04  &    2.65 \\
HD 34700     &      3.45  &    1.83 \\
HD 36112     &      1.75  &    2.08 \\
HD 245185    &      0.24  &    0.46  \\
HD 97048     &      1.81  &    0.57  \\
HD 100453    &      1.84  &    1.06 \\
HD 100546    &      1.00  &    0.27  \\
HD 135344    &      2.94  &    3.47 \\
HD 139614    &      1.65  &    0.88  \\
HD 142527    &      2.51  &    1.81 \\
CD$-$42$^\circ$11721  &      3.05  &    0.25  \\
HD 169142    &      2.50  &    1.00 \\
T Cra        &      2.28  &    0.42  \\
HD 179218    &      0.27  &    0.59  \\
HD 200775    &      2.83  &    0.88  \\
MWC 1080     &      1.98  &    1.79 \\
\hline
\textbf{Group II} &   &	  	   \\
VX Cas       &   $-$0.12  &     0.85  \\
HD 31648     &      0.09  &     2.51 \\
HD 35929     &   $-$1.02  &    21.56 \\
HD 244604    &   $-$0.67  &     2.73 \\
V586 Ori     &   $-$0.29  &     2.06 \\
BF Ori       &      0.47  &     3.23 \\
HD 95881     &   $-$1.97  &     2.69 \\
HD 98922     &   $-$2.03  &     2.42 \\
HD 101412    &   $-$0.70  &     1.80 \\
HD 104237    &   $-$0.49  &     3.55 \\
HD 141569    &      2.50  &     8.99 \\
HD 142666    &   $-$0.16  &     1.73 \\
HD 144432    &   $-$0.29  &     2.28 \\
HR 5999      &   $-$0.85  &     5.40 \\
HD 150193    &   $-$0.96  &     1.63 \\
AK Sco       &      0.85  &     2.63 \\
HD 163296    &      0.41  &     2.42 \\
VV Ser       &   $-$0.39  &     2.80 \\
WW Vul       &   $-$0.10  &     2.38 \\
HD 190073    &   $-$1.44  &     3.37 \\
BD$+$46$^\circ$3471   &      0.27  &     5.39 \\
SV Cep       &   $-$0.39  &     0.68  \\
IL Cep       &   $<$0.71  &  $>$3.16 \\
\hline
\textbf{Group III} &   &  		   \\
Z CMa        &      0.87  &    0.95 \\
MWC 297      &      1.90  &    1.05 \\
LkH$\alpha$ 225 &      1.66  &    0.03  \\
PV Cep       &      1.48  &    0.37 \\
V645 Cyg     &      1.27  &    0.33 \\
\hline
\end{tabular}
\end{center}
\end{table}

\begin{figure*}[!thp]
\resizebox{\textwidth}{!}{\rotatebox{0}{\includegraphics{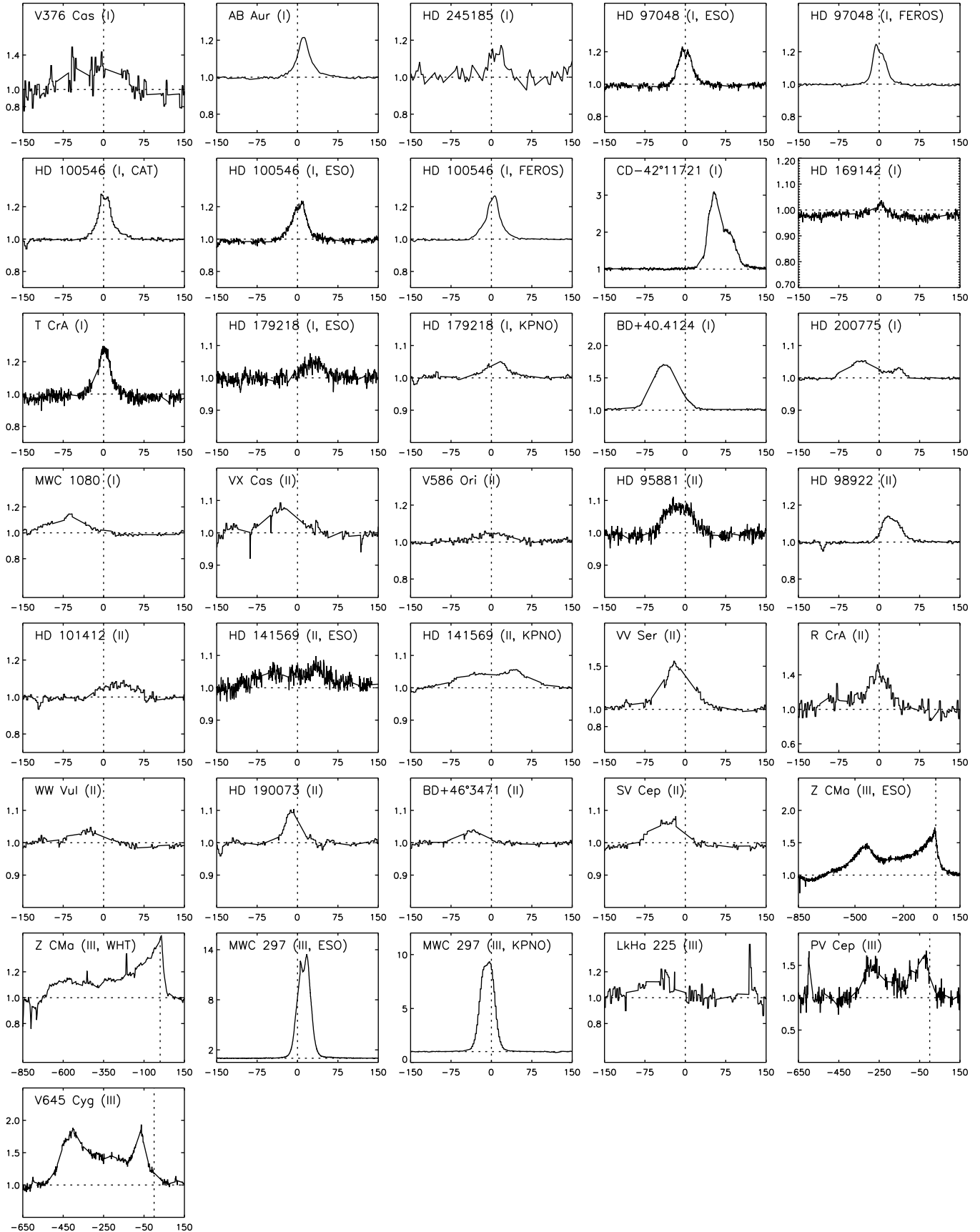}}} 
\caption{ The detected \OI 6300\A emission line spectra. Along the x-axis, the velocity
  in \kms~is given, on the y-axis the normalized intensity. The horizontal
  dotted line indicates the continuum 
  level, the vertical line the stellar radial velocity ($v_{rad} =
  0$). The source's name and group (in parentheses) are indicated in
  the upper left corner of each frame. When more than one spectrum is
  available, the type of spectrum is included as well.}
\label{specE.ps}
\end{figure*}

\begin{figure*}[!thp]
\resizebox{\textwidth}{!}{\rotatebox{0}{\includegraphics{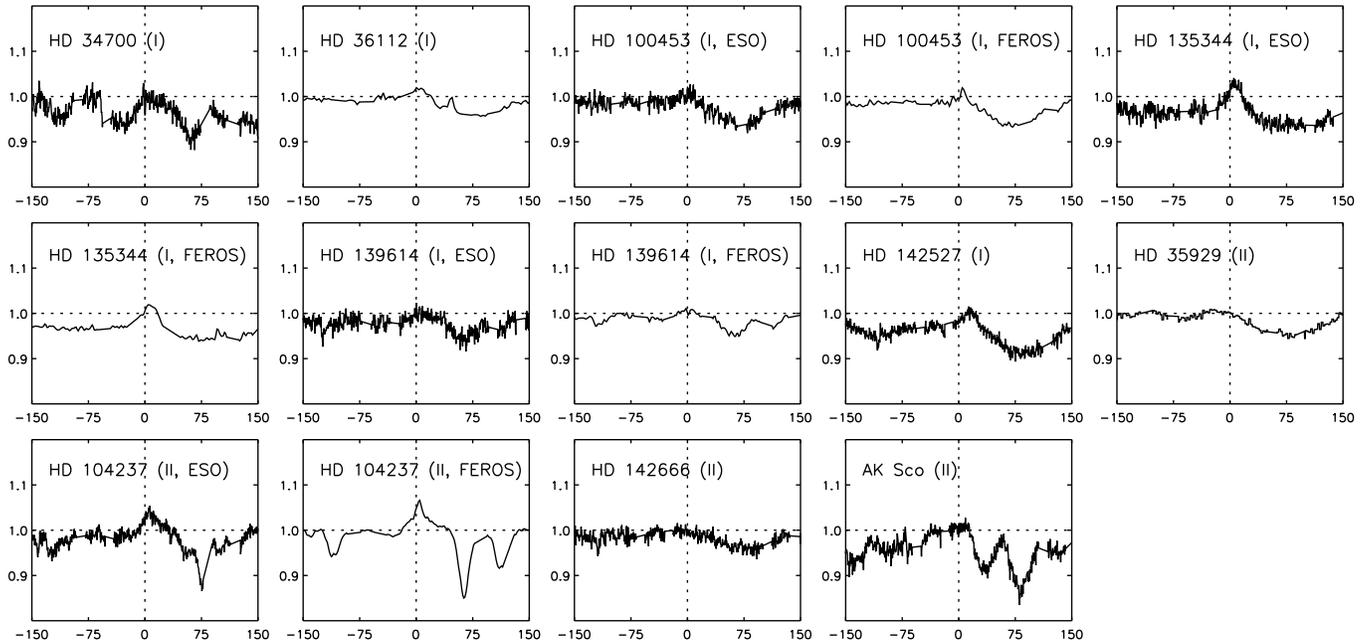}}} 
\caption{ The 6300\A spectra of the sample stars for which the
  detection of \OI emission is confused by photospheric absorption
  lines. Legend see Fig.~\ref{specE.ps}.}
\label{specA.ps}
\end{figure*}

\begin{figure*}[!thp]
\resizebox{\textwidth}{!}{\rotatebox{0}{\includegraphics{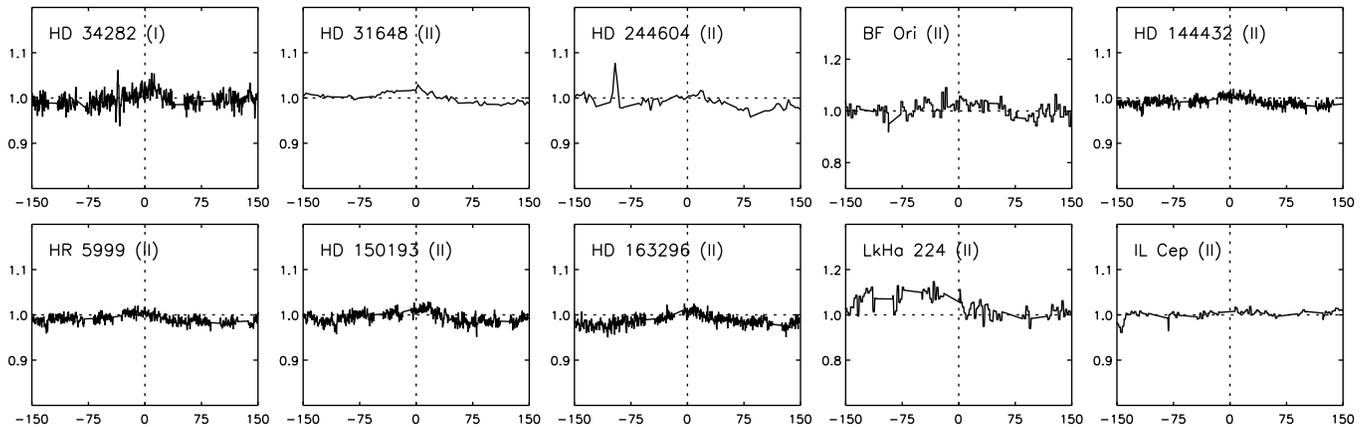}}} 
\caption{The 6300\A spectra of the sample stars with undetected \OI
  emission. Legend see Fig.~\ref{specE.ps}.}
\label{specU.ps}
\end{figure*}

In Figs.~\ref{specE.ps}, \ref{specA.ps} and \ref{specU.ps}, the velocity-rebinned
spectra of the \OI 6300\A emission lines are presented. The first plot
contains the detected emission profiles. Fig.~\ref{specA.ps} shows the
spectra in which the possible detection of \OI emission is confused by
underlying photospheric absorption lines. Sample sources with
undetected 6300\A emission are displayed in Fig.~\ref{specU.ps}. In
each plot, the targets are listed according to group.

%-----------------------------------------------------------------

\section{Analysis of the emission lines \label{secanalysis}}

\subsection{The \OI emission lines \label{secanaloi}}

\begin{figure}
\rotatebox{0}{\resizebox{3.5in}{!}{\includegraphics{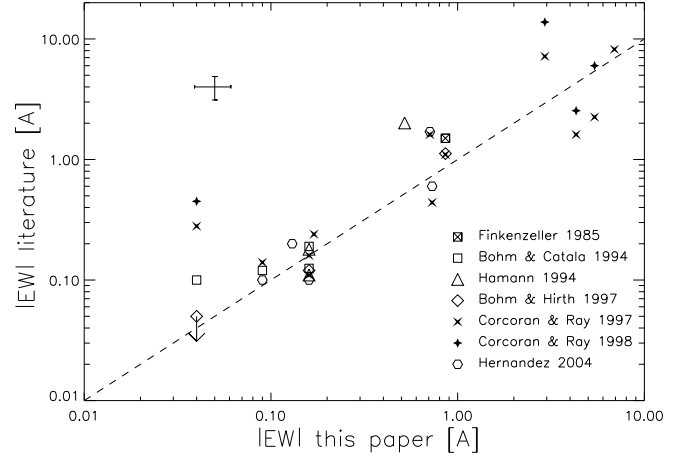}}} 
\caption{ Comparison between published values for the equivalent width
of the \OI 6300\A line and our determinations. The different plotting
symbols refer to data obtained by different authors. The typical error
bar is plotted 
in the upper left corner. The error bar on the literature values is
expected to be larger than the indicated error, which is equal to our
estimate for the error bar. The dashed line represents $\Delta$ EW = 0.}
\label{EWvslit.ps}
\end{figure}

The forbidden oxygen 6300\A emission line in Herbig Ae/Be stars has
been studied in previous articles
\citep{finkenzeller85,bohm94,hamann94,bohm97,corcoran97,corcoran98,hernandez04}. 
We have compared the equivalent widths measured in our
spectra to published literature values in Fig.~\ref{EWvslit.ps}. The error on
our EWs is computed from the noise on the data, multiplied with the
average FWHM of the feature. Few previous authors indicate the errors on their
measurements, hence we cannot set an error bar on the literature
numbers. Nevertheless we are convinced that our data set, consisting of
high-resolution, high-signal-to-noise spectra (S/N$\sim$30--375), are
of better quality than the data studied in previous papers. We
believe that our error bars are a reliable estimate for the scatter on
our determinations, and a lower limit to the literature errors.
The observed scatter in Fig.~\ref{EWvslit.ps} is
quite large. This might be due to intrinsicly variable \OI emission
\citep[e.g. \object{Z CMa},][]{vandenancker04}
or variable continuum emission (e.g. in UX Orionis
stars). A possible extrinsic explanation is the uncorrected airglow
emission in previous articles. Since we have removed the telluric
lines and airglow wavelength regions from our data, our measurements
are less affected by these features. Low spectral resolution does not
allow for removal of the latter. Most of the literature EW values in
Fig.~\ref{EWvslit.ps} --especially those of the weak lines-- appear to
be larger than our values, suggesting that the airglow effect indeed
corrupts these measurements.

In Section~\ref{measoi}, we have categorized the profiles based on the
presence of absorption features underlying the \OI emission line at
6300\r{A}. The latter features are photosperic absorption
lines, which have no direct relation with the \OI emission. This idea
is supported by the fact that all 10 sources which display absorption
features in the 6300\A region have late spectral types (i.e. low
effective temperatures; $6250\mathrm{K} < T_{\eff} <
8400\mathrm{K}$). Such 
objects display photosperic absorption lines in this
wavelength region, contrary to early-type stars which have flat
continua around 6300\r{A}.
 
\begin{figure*}
\rotatebox{0}{\resizebox{\textwidth}{!}{\includegraphics{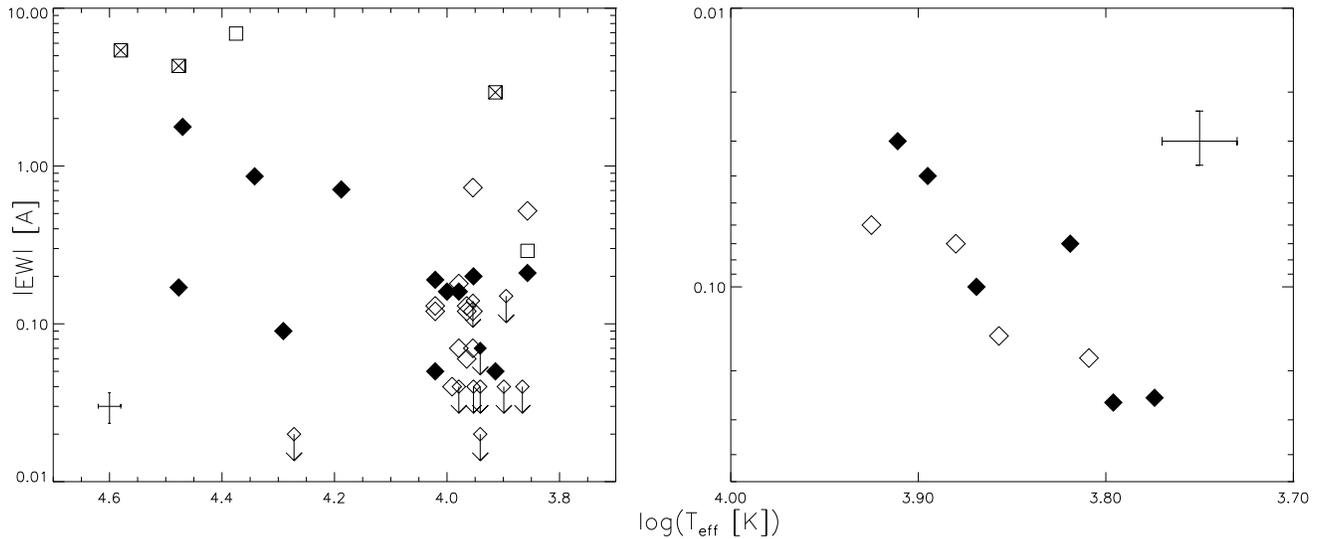}}} 
\caption{ The EWs of the \OI emission lines versus the effective
  temperature of the central star. Filled diamonds refer to group~I
  sources, open diamonds to 
  group~II. Squares are group~III sources, with the x-marks indicating
  the three sources with an extended highly-blueshifted wing. On the
  y-axis, $|$EW$|$ is plotted, on the x-axis the logarithm of the 
  effective temperature. The typical error bar is indicated in the
  upper right corner. The right panel contains the 10 sample stars
  for which the 6300\A region is rich in photospheric
  lines. The EWs are positive due to dominance of the absorption
  spectrum. The y-axis is flipped to make clear that increasing EW
  designates an increasing contribution of the absorption features.
  The other 39 sample sources (of which 29 with detected \OI 6300\A
  emission) are plotted in the left panel. These objects have earlier
  spectral types.
} 
\label{TeffvsOI.ps}
\end{figure*}

In Fig.~\ref{TeffvsOI.ps}, the measured EWs of the line
profiles is plotted versus the effective temperature. In the left part
of the figure, the targets with detected \OI emission are
plotted. These appear to be the early-type sources.
For stars in which the possible \OI 6300\A emission is confused by
photospheric lines (right part of the plot), the EWs increase
with decreasing $T_{\eff}$. This is probably the combined effect of
the increasing strength of the photospheric absorption lines on one
hand, and the decreasing UV luminosity in late-type stars on the
other (see later). Because of the difficulty to distinguish between
the possible \OI
emission line and the underlying photospheric absorption spectrum, no
accurate measurements of the FWHM and centroid position could be
made for these objects. Therefore, these determinations have been left
out of the further analysis. 

\begin{figure}
\rotatebox{0}{\resizebox{3.5in}{!}{\includegraphics{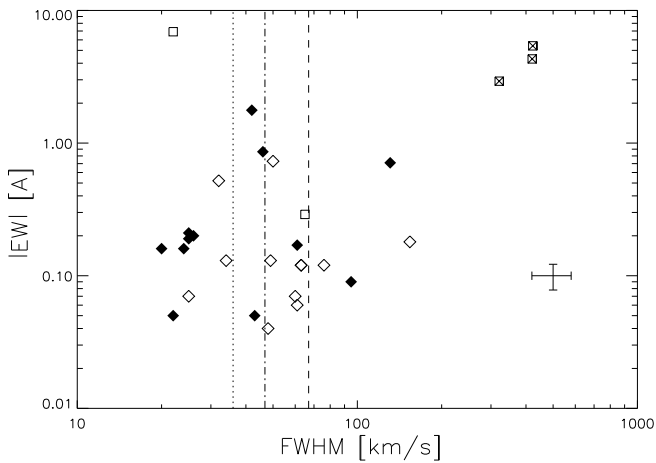}}} 
\caption{ The EW of the detected \OI emission lines versus the FWHM. Filled
  diamonds represent group~I, open diamonds group~II and
  squares group~III sources. The x-marked squares represent \hbox{PV
  Cep}, \hbox{V645 Cyg} and \hbox{Z CMa}. The dash-dotted line indicates
  the mean FWHM (47 \kms). The dotted and dashed lines indicate the
  mean values of group~I (34 \kms) and group~II (67 \kms) respectively.
  The typical error bar is represented on the
  right side of the plot.}
\label{EWvsFWHM.ps}
\end{figure}

\begin{figure}
\rotatebox{0}{\resizebox{3.5in}{!}{\includegraphics{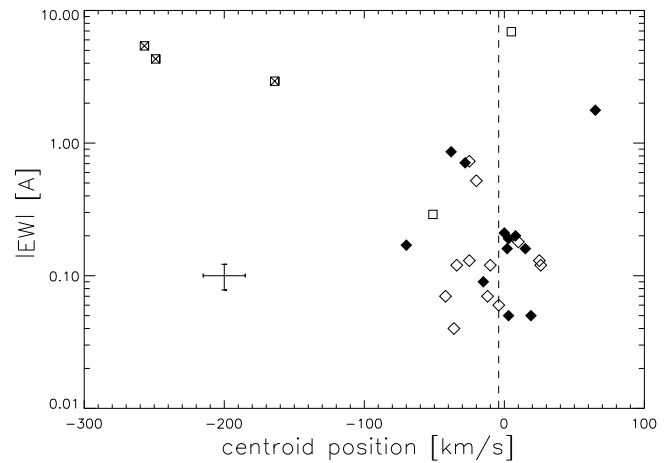}}} 
\caption{ The EW of the detected \OI emission lines versus the centroid
  position of the feature. The plotting symbols are as in
  Fig.~\ref{EWvsFWHM.ps}. The vast majority of the features is
  centered around the stellar radial velocity within the error
  bars. The dashed line represents the average centroid position ($-$4
  \kms). The typical error bar is represented in the lower left corner.}
\label{EWvscentroid.ps}
\end{figure}

\begin{figure}
\rotatebox{0}{\resizebox{3.5in}{!}{\includegraphics{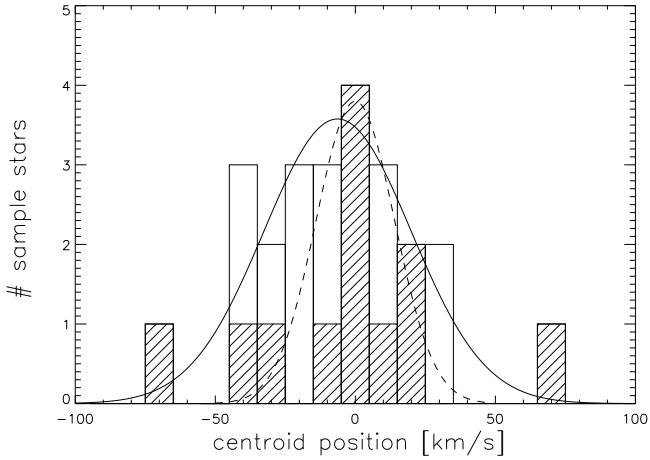}}}
\caption{ The histogram of the centroid positions of the 12 group~I and
  12 group~II sample sources with a detected \OI 6300\A feature. The bin
  size is 10 \kms. The  
  dashed line represents distribution $D_{\mathrm{exp}}$, the full line
  is the Gaussian fit to the observed distribution. The cross-hatched
  part of the bars indicate 
  group~I sources, the open parts group~II members.}
\label{histo_centroidsmallvel.ps}
\end{figure}

\begin{figure}
\rotatebox{0}{\resizebox{3.5in}{!}{\includegraphics{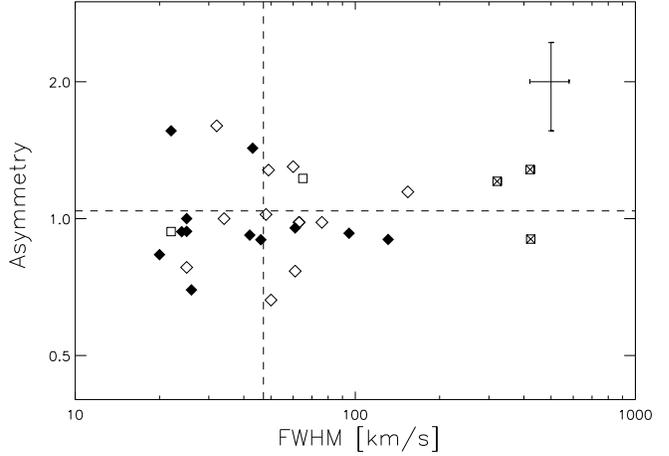}}} 
\caption{ The FWHM of the detected \OI emission lines versus the asymmetry
  parameter of the feature. The plotting symbols are as in 
  Fig.~\ref{EWvsFWHM.ps}. The dashed horizontal line represents the mean value (1.04)
  for $\mathcal{A}$, the vertical line indicates the mean FWHM (47
  \kms). The typical error bar is represented in the 
  upper right corner.}
\label{FWHMvsasym.ps}
\end{figure}

The other sample sources display no absorption lines in the \OI
wavelength range. 10 objects do not display any significant
features and will be included in the figures as upper limits. The
majority of our sample sources (29/49) have a clear \OI emission
profile in their spectra. In
Figs.~\ref{EWvsFWHM.ps} to \ref{FWHMvsasym.ps}, the parameters that
describe the emission \OI profile are plotted against each
other. The mean 
parameters describing the emission line profile are
$\langle$FWHM$\rangle = 47$ \kms, $\langle$centroid position$\rangle =
-$4 \kms\ and $\langle \mathcal{A} \rangle = 1.04$. The median
  values of these quantities in this sample
are 47 \kms, -7 \kms and 0.95 respectively. It is striking
that, when excluding the three outliers (\object{PV Cep}, \object{V645
  Cyg} and \object{Z CMa}; $\boxtimes$ in the plots) which will
be discussed later, most of
the sample sources with pure-emission \OI profiles lie 
close to these average values. The objects seem to have quite
uniform \OI emission line profiles. The EW of the feature
is not correlated to its width (FWHM), indicating that the broadening
mechanism of the line does not depend on the amount of \OI
emission. Nevertheless, a difference in FWHM is observed between
group~I (mean: 36 \kms; median: 34 \kms) and group~II (mean: 67 \kms;
median: 55 \kms). The centroid position is also independent of the EW
(Fig~\ref{EWvscentroid.ps}).

\citet{bohm94} note that the \OI 6300\A
profiles in their sample are not blueshifted with respect to the
stellar radial velocity. \citet{corcoran97} on the other hand report
that most of the HAEBEs in their sample show low-velocity blueshifted
centroid positions. We look into this more carefully.
Fig.~\ref{histo_centroidsmallvel.ps} displays the distribution of the centroid
positions of the pure-emission \OI profiles in a histogram. The group
III sources have been excluded from the diagram. The width of the
bins is 10 \kms. 
The estimated error on the measured centroid positions,
including the error on the stellar-radial-velocity determination, is
$\sigma = 15$ \kms. We assume that the errors on the centroid
positions are distributed following a standard-normal distribution
$D_{\mathrm{exp}}$.
We check whether the observed distribution $D_{\mathrm{obs}}$ is 
representative for the expected normal distribution
$D_{\mathrm{exp}}$. The possibility that the centroid position of a
feature is $>\sigma$ under the hypothesis of the
$D_{\mathrm{exp}}$ distribution is 16\%. Five sources display
redshifted centroid 
positions larger than 15 \kms. This is not a statistically significant deviation
from $D_{\mathrm{exp}}$ for a sample of 24 stars. Nevertheless, on the
blue side, nine sources display centroid positions $<-15$ \kms. The
possibility for this to happen under the current hypothesis is less
than 1\%. We conclude that, although the majority of stars in our
sample have centroid positions that are compatible with the stellar
radial velocity, there appear to be some objects that indeed show
low-velocity blueshifted centroid positions as reported by
\citet{corcoran97}. The Gaussian fit (full line in
Fig.~\ref{histo_centroidsmallvel.ps}) to the observed  
histogram is centered around $-$7 \kms\ and has $\sigma = 27$ \kms.
Notice that for the five sources with detected \OI 6300\A emission for
which we 
have spectra covering a large part of the optical wavelength region
(i.e. \object{AB Aur}, \object{HD 245185}, \object{HD 97048},
\object{HD 100546} and \object{Z CMa}), the error on the centroid
determination is much smaller than 15 km/s. Of these objects, only
group~III member \object{Z CMa} shows evidence for a blueshifted
feature. In particular the small error ($\sim$2 \kms) for \object{HD
100546} proves that at least some sources are definitely \textit{not}
blueshifted.

Even though some sources display blueshifted emission, there does not
seem to be a strong deviation from asymmetry. In
Fig.~\ref{FWHMvsasym.ps} the FWHM of the \OI feature is 
plotted versus the asymmetry parameter $\mathcal{A}$. No correlation
between the two parameters is noted.

Three group~III sources (\hbox{PV Cep}, \hbox{V645 Cyg} and \hbox{Z CMa})
have radically different \OI emission profiles from the other sources
in our sample. The shape of these
features is double-peaked, with one peak close to the radial velocity
of the central star ($v = 0$ \kms) and the other at high blueshift
($v \approx -350$ \kms). The red wing ($v > 0$ \kms) of the
feature extends up to $v \sim 50$ 
\kms, while the blue wing ($v < 0$ \kms) is much stronger and expands
as far as $v \sim -410$ to $-730$ \kms. This makes the \OI 6300\A line
in these sources
outliers in Figs.~\ref{EWvsFWHM.ps} and \ref{EWvscentroid.ps}.
Because of the obvious differences, the parameters describing the
\OI profiles of these three sources are not included in the
computation of the average \OI parameters. The asymmetry parameter on
the other hand is close to 1 for all three sources. Indeed, the
features are quite symmetric around the (strongly blueshifted)
centroid position. This is due to the fact that the blueshifted
emission peak in these features is of comparable strength to the
emission peak near 0 \kms.

For 30 objects, the 6363\A line region is covered by our spectra. In
15 sample sources, where the 
\OI 6300\A line was detected, the \OI 6363\A line was detected as
well. The profile of the \OI line in these objects is
pure-emission. Six sources for which the \OI 6300\A line was detected,
did not display the 6363\A line. In the remaining 9 objects, no \OI
lines were detected. This includes all 5 sources with an
absorption-line-rich 6300\A wavelength region, for which we have data
around 6363\r{A}. The mean ratio of the EWs of the 6300\A and 6363\A lines
$\mathrm{EW}(6300)/\mathrm{EW}(6363)$ is 3.1$\pm$0.9, which is within 
the error equal to the ratio of the Einstein transition rates
of the lines ($A_{6300}/A_{6363} = 3.0$). The lower limits derived
for the sources with an undetected 6363\A line are also consistent with
this value. The mean difference between the FWHMs of both lines is
close to 0 \kms, FWHM(6300\r{A})$-$FWHM(6363\r{A}) = 3$\pm$20 \kms.
The \OI 6363\A emission line has the same origin as
the \OI 6300\A line, as one would expect for two lines that
originate from the same upper energy level (${^1D_2}$). The spectral
information in one line is exactly the same as in the other. The only
parameter that can influence the EW ratio of both lines is
\textit{differential} extinction. Using a standard interstellar
extinction law \citep{fluks} and the error bar on the EW ratio
($\sim$30\%), one can derive that the visual
extinction towards the emission region must be less than 30.
Also for \object{Z CMa} ---in which the profile is very broad, and we
have a sufficient S/N in both the 6300 and 6363\A line to allow an
accurate comparison--- the two \OI lines have comparable profiles.
The [\SII] emission line at 6731\A --which is
present in the WHT spectrum of \object{Z CMa}-- has a
similar shape as the \OI lines, be it with a much weaker blue wing.
This forbidden line is undetected ($<0.01$\r{A}) in the other FEROS
and WHT spectra.

The wavelength region around the forbidden oxygen line at 5577\A is
covered by the FEROS and WHT spectra. No features where detected, with
upper limits of 0.01\r{A}. Assuming that the emission of
both the 5577\A and 6300\A line are thermal, one can estimate the
theoretical Saha-Boltzmann 5577/6300 intensity ratio. In our
sample we measure upper limits for this ratio of the order of
$\sim$1/20. The observations hence exclude that the oxygen lines are
produced by thermal emission of oxygen atoms at temperatures above
$\sim$3000K. This is a strong 
indication that the source of the \OI emission at 6300\A in HAEBEs
cannot be found in thermal excitation of oxygen in a ``super-heated''
surface layer, as is the case for T~Tauri stars (see later).

\begin{figure}
\rotatebox{0}{\resizebox{3.5in}{!}{\includegraphics{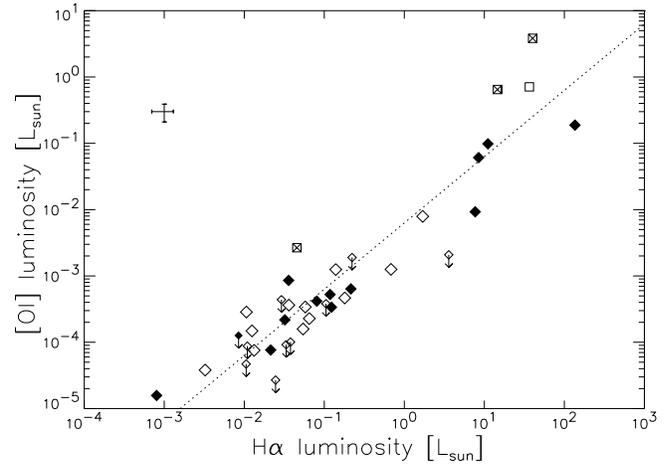}}} 
\caption{ The [O\,{\sc i}]-versus-\Ha luminosity. The sources with a
  negative \Ha EW have been included. The typical error
  bar is presented in the upper left corner. Plotting symbols are as
  in Fig.~\ref{EWvsFWHM.ps}. The dotted line represents the median
  luminosity ratio $L$([O\,{\sc i}])$/L(\mathrm{H}\alpha) = 6.3 \times
  10^{-3}$.}
\label{LhalphavsLOI.ps}
\end{figure}

The \Ha features in our sample were classified based on the observed
profile: single-peaked, double-peaked, P Cyg-like or inversed P
Cyg-like. Despite the object's name, in our spectrum of
\object{LkH$\alpha$ 225} no \Ha feature was detected. Since this
spectral feature has previously been observed in emission in both
components of this binary system \citep{magakyan97}, the absence of the line
in our spectrum suggests \Ha variability.
10 of the 49 sample sources (20\%) display single-peaked, 25 (51\%)
double-peaked and 12 (24\%) P Cyg-like H$\alpha$ emission
profiles. In \object{SV Cep}, an inversed P 
Cyg-profile is observed. This is in good agreement with the \Ha
distribution in the samples of \citet{finkenzeller84} and
\citet{bohm94}: 50\% double-peaked, 25\% single-peaked and 20\% P
Cyg-like. In Table~\ref{halphavsoi} the different categories of the
\Ha profile are listed versus the types of the \OI profile.
When plotting the \OI luminosities of the sample stars versus the \Ha
luminosities, regardless of the type of profile, a clear correlation
is noted (Fig.~\ref{LhalphavsLOI.ps}). \object{Z CMa}, \object{PV Cep}
and \object{V645 Cyg} seem to have significantly more [O\,{\sc i}]-to-\Ha
luminosity than other sources. No significant differences 
between the other groups are noted. The average luminosity ratio
$L$([O\,{\sc i}])$/L(\mathrm{H}\alpha)$ is $1.4 \times 10^{-2}$.

\begin{table}
\caption{ The different categories of \Ha profiles versus the types
  of \OI profiles. The entries of the table are the number of
  stars belonging to the corresponding category. U: undetected
  feature, E: pure-emission profile, A: 6300\A region rich in
  photospheric absorption lines, S: single-peaked profile, D: double-peaked
  profile, P: \hbox{P Cyg}-like profile, inv. P: inversed \hbox{P
  Cyg}-like profile. 
\label{halphavsoi}}
\begin{center}
\begin{tabular}{l|ccc|c}
 \multicolumn{1}{r|}{\OI} &
 \multicolumn{1}{|c}{U} &
 \multicolumn{1}{c}{E} &
 \multicolumn{1}{c|}{A} &
 \multicolumn{1}{|c}{Tot.} \\
 \multicolumn{1}{l|}{\Ha~~~~~~~~~} &
 \multicolumn{3}{|c|}{} & 
 \multicolumn{1}{|c}{}\\
 \hline
U       & 0 & 1 & 0 & 1  \\
S       & 0 & 5 & 5 & 10 \\
D       & 6 & 15& 4 & 25 \\
P       & 4 & 7 & 1 & 12 \\
inv. P  & 0 & 1 & 0 & 1  \\
 \hline
Tot.    &10 & 29& 10& 49 \\
\end{tabular}
\end{center}
\end{table}

\subsection{\OI versus the SED parameters \label{oivssed}}

\begin{figure}
\rotatebox{0}{\resizebox{3.5in}{!}{\includegraphics{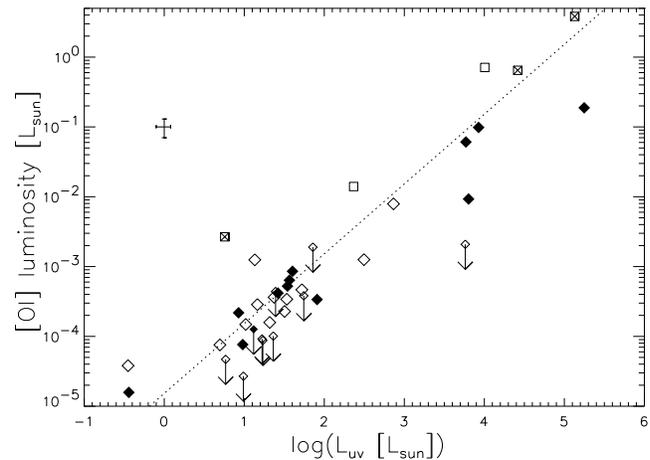}}} 
\caption{ The \OI 6300\A luminosity versus the stellar UV
  luminosity. The plotting symbols are as in
  Fig.~\ref{EWvsFWHM.ps}. The typical error bar is indicated in the
  upper left corner. The dotted line represents the median luminosity
  ratio $L$([O\,{\sc i}])$/L_{\uv} = 1.5 \times 10^{-5}$. }
\label{LUVvsLOI.ps}
\end{figure}

In this section we compare the parameters describing the \OI 6300\A
emission line to the SED parameters. In Fig.~\ref{LUVvsLOI.ps} the \OI
luminosity is plotted versus the UV luminosity of the central
star. The dashed line represents the median [O\,{\sc i}]-over-UV
luminosity ratio ($1.5 \times 10^{-5}$) of the detected lines, while
the average 
value is $\sim$3.9 $\times 10^{-5}$. In general, the \OI luminosity
increases with increasing $L_{\uv}$. The different plotting symbols
represent the three groups of HAEBE stars in our sample.
Note that there is a considerable number of (almost all group~II)
sources with an undetected \OI 6300\A feature. The upper limits
clearly deviate from the average.

\begin{figure}[!thp]
\resizebox{3.5in}{!}{\rotatebox{0}{\includegraphics{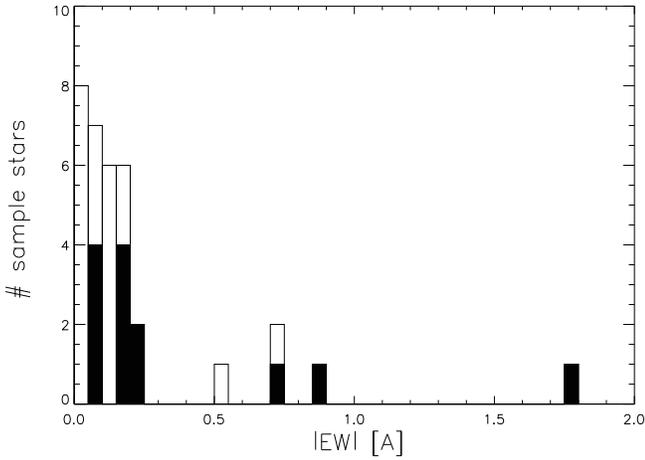}}} 
\caption{ Histogram of the EWs containing the 13 group~I and 24 group~II
  members for which we have a measurement of, or an upper limit to the
  EW of the \OI 6300\A emission line. The filled part
  of the bars represents group~I sources, the open part group~II
  members. The bin size is 0.05\r{A}. The group~I sources have
  slightly higher EW values than their group~II counterparts.}
\label{histogramEW.ps}
\end{figure}

\begin{figure}[!thp]
\resizebox{3.5in}{!}{\rotatebox{0}{\includegraphics{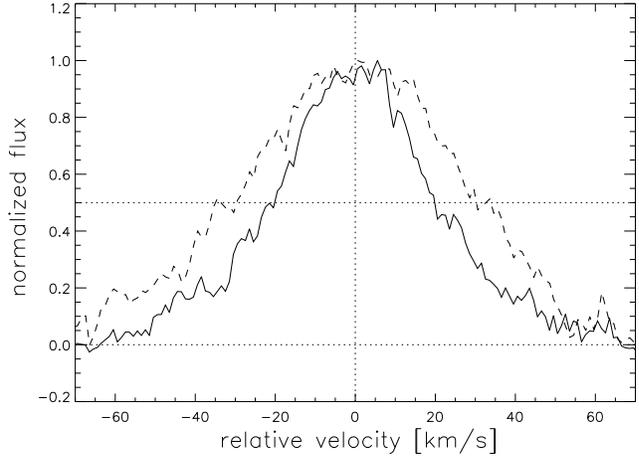}}} 
\caption{ The average \OI 6300\A profiles for
  group~I and group~II. The horizontal dotted lines are plotted at 0 and
  0.5, while the vertical dotted line represents the centroid position
  of the feature. The 12 group~I and 12 group~II sources which
  display \OI emission are included. The profile of each source was
  shifted to its centroid position, continuum-subtracted and
  normalized by dividing by its peak-flux. The group~II profile
  (dashed line) is  
  $\sim$20 \kms broader than the group~I profile (full line).}
\label{profileIvsIIBW.ps}
\end{figure}

The group~III sources display the strongest \OI
emission. All 5 sources in this group have high $L$([O\,{\sc i}]) in the range
between $2.7 \times 10^{-3}$ and $3.8 L_\odot$. The median \OI 6300\A
luminosity is $6.5 \times 10^{-1} L_\odot$. Of the 12 group~I objects
in the plot, only one source does  
not show a detected \OI 6300\A emission line. The median \OI luminosity of
the detected features in this group is $5.9 \times 10^{-4}
L_\odot$. About 43\% of the group~II sources in this sample (excluding
those with absorption-polluted spectra) do not 
have a detected \OI emission line. Moreover, even the median
luminosity of the \textit{detected} features in group~II ($3.2 \times
10^{-4} L_\odot$) is smaller than in the other two groups.
Fig.~\ref{histogramEW.ps} shows the histogram of the EWs of the group
I and II sources. The group~I sources have slightly stronger \OI
intensities than group~II sources. Also the typical width of the features
differs between group~I and II. The latter is illustrated in
Fig.~\ref{profileIvsIIBW.ps}. The group~II profile is $\sim$20 \kms
broader than the group~I profile.
We conclude that there is a clear correlation between the IR-SED
classification of the sources and the strength and shape of the \OI
emission line.

In Table~\ref{samplevsOI} we have summarized the parameters
describing the \OI emission lines. The sample sources are listed
according to their classification.

\begin{table*}[!th]
\caption{ The \OI parameters for each sample source. The objects are
  listed according to their IR-SED classification. The distance to the
  source, the effective temperature, UV luminosity and bolometric
  (= integrated \textit{observed}) luminosity are given. References
  for the first two values can be found in Table 6 of AV04, unless
  otherwise indicated: $^A$ \citet{dezeeuw}; $^B$
  \citet{finkenzeller85}; $^C$ \citet{gray93}; $^D$ \citet{gray};
  $^E$ \citet{houk78}; $^F$ \citet{mora01}; $^G$ \citet{vieira03};
  $^H$ \citet{vandenancker98}. The type of \OI
  6300\A emission profile observed in each object has been indicated;
  E: pure-emission profile, A: 6300\A region rich in photospheric
  absorption lines, U: undetected feature (upper limit). n.s.: no spectrum
  available. The typical errors on the FWHM and centroid position are
  $\pm8$ and $\pm15$ \kms\, respectively.
  The thirth column contains the stellar radial velocity employed in
  this study. This value is estimated from the 
  available spectra, unless otherwise indicated: $^a$ \citet{arce};
  $^b$ \citet{arellanoferro}; $^c$ \citet{barbierbrossat}; $^d$
  \citet{dunkin97a}; $^e$ \citet{pietu}; $^f$ \citet{reipurth}; $^g$
  \citet[][based on CO lines]{yonekura}. $^\dagger$ Based on Ca {\sc ii}
  lines. $\circlearrowright$ Average of radial velocities of
  stars in the same field (radius$\sim$2\arcmin). 
  The last two columns of the table give the type and EW of the
  observed \Ha profile; S: single-peaked; D: double-peaked; P: P
  Cygni-like; inv. P: inversed P Cyg-like. When no spectrum was
  available, literature 
  values were used: $^h$ \citet{arellanoferro}; $^i$ \citet{dewinter};
  $^j$ \citet{dunkin97a}; $^k$ \citet{dunkin97b}; $^l$
  \citet{finkenzeller84}; $^m$ \citet{merin04}; $^n$ \citet{perez}.
\label{samplevsOI}}
\begin{center}
\scriptsize
\tabcolsep0.11 cm
\begin{tabular}{cccccc|ccccc|cc|cc}
 \multicolumn{6}{c|}{} &
 \multicolumn{5}{|c|}{\OI 6300\A} &
 \multicolumn{2}{|c|}{\OI 6363\A} & 
 \multicolumn{2}{|c}{\Ha}\\
 \multicolumn{1}{c}{Object} &
 \multicolumn{1}{c}{Distance} &
 \multicolumn{1}{c}{$\log T_\eff$} &
 \multicolumn{1}{c}{$\log L_\uv$} &
 \multicolumn{1}{c}{$\log L_\bol$} &
 \multicolumn{1}{c|}{$v_{rad}$} &
 \multicolumn{1}{|c}{Profile} &
 \multicolumn{1}{c}{EW} &  
 \multicolumn{1}{c}{$L$} &
 \multicolumn{1}{c}{FWHM} &
 \multicolumn{1}{c|}{Centroid} &
 \multicolumn{1}{|c}{EW} &
 \multicolumn{1}{c|}{FWHM} &
 \multicolumn{1}{|c}{Profile} &
 \multicolumn{1}{c}{EW} \\ 
 \multicolumn{1}{c}{} &
 \multicolumn{1}{c}{[pc]} &
 \multicolumn{1}{c}{$\log$ [K]} &
 \multicolumn{1}{c}{$\log\ [L_\odot]$} &
 \multicolumn{1}{c}{$\log\ [L_\odot]$} &
 \multicolumn{1}{c|}{[\kms]} &
 \multicolumn{1}{|c}{} &
 \multicolumn{1}{c}{[\r{A}]} &
 \multicolumn{1}{c}{[$L_\odot$]} &
 \multicolumn{1}{c}{[\kms]} &
 \multicolumn{1}{c|}{[\kms]} &
 \multicolumn{1}{|c}{[\r{A}]} &
 \multicolumn{1}{c|}{[\kms]} &
 \multicolumn{1}{|c}{} &
 \multicolumn{1}{c}{[\r{A}]} \\
\hline
\textbf{Group I} & & & & & & & & & & & & & & \\
V376 Cas             &  630     & 4.19    &    1.60  &  2.58   & $-$28$^g$                 & E &   $-$0.7$\pm$0.2   & $8.6 \times 10^{-4}$  &   131  & $-$28 &$|$EW$|$ $<$ 0.7  &  &  D  & $-$29.7   \\
AB Aur               &  144     & 3.98    &    1.56  &  1.71   &   9                       & E &  $-$0.16$\pm$0.01  & $6.4 \times 10^{-4}$  &    20  &    15 &$-$0.04  &  20	 &  P  & $-$54.0   \\
HD 34282             &  400     & 3.94    &    1.12  &  1.38   &  16$^e$                   & U & $|$EW$|$ $<$ 0.07  & $< 1.3 \times 10^{-4}$&        &       &   n.s.   &  n.s.	 &  D  & $-$4.7$^m$ \\
HD 34700             &  336     & 3.77    &    0.91  &  1.42   & $-$21$^b$                 & A &     0.25$\pm$0.05  &                       &        &       &   n.s.   &  n.s.	 &  S  &  $-$0.6$^h$ \\
HD 36112             &  204$^H$ & 3.91$^D$&    1.16  &  1.47   &  12                       & A &     0.03$\pm$0.01  &                       &        &       &$|$EW$|$ $<$ 0.02 &  &  P  & $-$19.7   \\
HD 245185            &  336$^A$ & 3.95$^D$&    0.93  &  1.29   &  24                       & E &  $-$0.20$\pm$0.08  & $2.2 \times 10^{-4}$  &    26  &     8 &$|$EW$|$ $<$ 0.1 &   &  D  & $-$29.5   \\
HD 97048             &  180     & 4.00    &    1.54  &  1.42   &  21                       & E &  $-$0.16$\pm$0.01  & $5.3 \times 10^{-4}$  &    24  &     2 &$-$0.05  &  21	 &  D  & $-$36.0   \\
HD 100453            &  112     & 3.87    &    0.66  &  1.09   &  17                       & A &     0.10$\pm$0.01  &                       &        &       &$|$EW$|$ $<$ 0.02 &  &  D  &   1.5   \\
HD 100546            &  103     & 4.02    &    1.42  &  1.62   &  18                       & E &  $-$0.19$\pm$0.01  & $4.2 \times 10^{-4}$  &    25  &     3 &$-$0.06  &  21	 &  S  & $-$36.8   \\
HD 135344            &  140     & 3.82    &    0.60  &  1.01   &  $-$1                     & A &     0.07$\pm$0.01  &                       &        &       &$|$EW$|$ $<$ 0.02 &  &  S  & $-$17.4   \\
HD 139614            &  140     & 3.89    &    0.71  &  1.03   &   5                       & A &     0.04$\pm$0.01  &                       &        &       &$|$EW$|$ $<$ 0.02 &  &  S  &  $-$9.8   \\
HD 142527            &  145     & 3.80    &    0.83  &  1.34   &  $\circlearrowright{-9}$  & A &     0.26$\pm$0.02  &                       &        &       &   n.s.   &  n.s.	 &  S  & $-$17.9$^n$ \\
CD$-$42$^\circ$11721 &  400     & 4.47    &    3.93  &  3.03   & $-$84$^f$                 & E &  $-$1.77$\pm$0.03  & $9.8 \times 10^{-2}$  &    42  &    65 &   n.s.   &  n.s.	 &  S  &$-$199.5$^i$ \\
HD 169142            &  145     & 3.91    &    0.98  &  1.13   &  $-$3$^d$                 & E &  $-$0.05$\pm$0.02  & $7.6 \times 10^{-5}$  &    22  &     3 &   n.s.   &  n.s.	 &  S  & $-$14.0$^k$ \\
T Cra                &  130     & 3.86    &   -0.44  &  0.88   &   0$^f$                   & E &  $-$0.21$\pm$0.04  & $1.6 \times 10^{-5}$  &    25  &     0 &   n.s.   &  n.s.	 &  D  &  $-$10.7   \\
HD 179218            &  240     & 4.02    &    1.91  &  1.88   &  $-$9                     & E &  $-$0.05$\pm$0.01  & $3.4 \times 10^{-4}$  &    43  &    19 &$-$0.03  &  29	 &  S  &  $-$18.2   \\
BD$+$40$^\circ$4124  &  980     & 4.34    &    3.77  &  2.50   &  14                       & E &  $-$0.86$\pm$0.01  & $6.1 \times 10^{-2}$  &    46  & $-$38 &$-$0.26   &  41	 &  D  & $-$119.9   \\
HD 200775            &  440     & 4.29    &    3.80  &  2.84   &   8                       & E &  $-$0.09$\pm$0.01  & $9.3 \times 10^{-3}$  &    95  & $-$15 &$-$0.04  &  107	 &  D  & $-$74.6   \\
MWC 1080             &  2200    & 4.48    &    5.25  &  3.76   &  $-$4                     & E &  $-$0.17$\pm$0.02  & $1.9 \times 10^{-1}$  &    61  & $-$70 &$-$0.03  &  42	 &  P  & $-$122.8   \\
\hline		       	     		   	      					    						       	       
\textbf{Group II} & & & & & & & & & & & & & & \\
VX Cas               &  630     & 3.97    &    1.37  &  1.45   &   7                       & E &  $-$0.13$\pm$0.02  & $3.6 \times 10^{-4}$  &    49  & $-$25 &$-$0.07  &  59	 &  D  &  $-$12.9   \\
HD 31648             &  131     & 3.94    &    0.99  &  1.22   &  11                       & U & $|$EW$|$ $<$ 0.02  & $< 2.7 \times 10^{-5}$&       &       &$|$EW$|$ $<$ 0.02 & &  P  & $-$18.3   \\
HD 35929             &  510$^A$ & 3.86$^D$&    1.81  &  1.95   &  15$^\dagger$             & A &     0.15$\pm$0.03  &                       &        &       &   n.s.   &  n.s.	 &  S  & $-$4.8   \\
HD 244604            &  336$^A$ & 3.98$^D$&    1.36  &  1.35   &  21                       & U & $|$EW$|$ $<$ 0.04  & $< 1.0 \times 10^{-4}$&       &       &$|$EW$|$ $<$ 0.03 & &  D  &  $-$15.0   \\
V586 Ori             &  510$^A$ & 3.97$^C$&    1.50  &  1.61   &  26$^f$                   & E &  $-$0.06$\pm$0.02  & $2.3 \times 10^{-4}$  &    61  &  $-$4 &   n.s.   &  n.s.	 &  D  &  $-$17.1   \\
BF Ori               &  510     & 3.95    &    1.39  &  1.38   &  26$^f$                   & U & $|$EW$|$ $<$ 0.14  & $< 4.4 \times 10^{-4}$&       &       &   n.s.   &  n.s.	 &  D  &  $-$9.3   \\
HD 95881             &  118     & 3.95    &    0.70  &  0.99   &  36$^\dagger$             & E &  $-$0.12$\pm$0.02  & $7.6 \times 10^{-5}$  &    76  & $-$10 &   n.s.   &  n.s.	 &  D  & $-$21.1   \\
HD 98922            &$>$540$^H$ & 4.02$^E$&    2.86  &  2.95   & $-$15$^\dagger$           & E &  $-$0.13$\pm$0.01  & $7.9 \times 10^{-3}$  &    34  &    25 &   n.s.   &  n.s.	 &  P  & $-$27.9   \\
HD 101412            &  160$^A$ & 4.02$^G$&    1.53  &  1.40   &  $\circlearrowright{-3}$  & E &  $-$0.12$\pm$0.03  & $3.4 \times 10^{-4}$  &    63  &    26 &   n.s.   &  n.s.	 &  D  & $-$20.4   \\
HD 104237            &  116     & 3.92    &    1.36  &  1.53   &  13                       & A &     0.06$\pm$0.01  &                       &        &       &$|$EW$|$ $<$ 0.02&   &  D  & $-$24.5   \\
HD 141569            &  99      & 3.98    &    1.16  &  1.10   &  $-$2                     & E &  $-$0.18$\pm$0.01  & $2.9 \times 10^{-4}$  &   154  &    10 &$-$0.09  &  158	 &  D  & $-$6.7   \\
HD 142666            &  145     & 3.88    &    0.91  &  1.03   &  3$^d$                    & A &     0.07$\pm$0.02  &                       &        &       &   n.s.   &  n.s.	 &  D  &  $-$3.2$^j$       \\
HD 144432            &  145     & 3.87    &    0.77  &  1.13   &  2$^d$                    & U & $|$EW$|$ $<$ 0.04  & $< 4.7 \times 10^{-5}$&       &       &   n.s.   &  n.s.	 &  P  &  $-$9.0$^j$       \\
HR 5999              &  210     & 3.90    &    1.74  &  1.98   &  16$^\dagger$             & U & $|$EW$|$ $<$ 0.04  & $< 3.8 \times 10^{-4}$&       &       &   n.s.   &  n.s.	 &  D  &  $-$11.0   \\
HD 150193            &  150     & 3.95    &    1.23  &  1.19   &  $-$6$^f$                 & U & $|$EW$|$ $<$ 0.04  & $< 8.8 \times 10^{-5}$&       &       &   n.s.   &  n.s.	 &  P  &  $-$5.0$^l$       \\
AK Sco               &  150     & 3.81    &    0.63  &  0.89   &  0$^f$                    & A &     0.18$\pm$0.03  &                       &        &       &   n.s.   &  n.s.	 &  D  &  $-$9.4   \\
HD 163296            &  122     & 3.94    &    1.22  &  1.52   &  $-$4$^c$                 & U & $|$EW$|$ $<$ 0.04  & $< 9.2 \times 10^{-5}$&       &       &   n.s.   &  n.s.	 &  D  & $-$14.5$^l$        \\
VV Ser               &  330     & 3.95    &    1.13  &  1.37   &  $-$4                     & E &  $-$0.73$\pm$0.04  & $1.2 \times 10^{-3}$  &    50  & $-$25 &$-$0.21   &  47	 &  D  & $-$81.3   \\
R CrA                &  130     & 3.86    &   -0.45  &  3.46   &  0$^f$                    & E &  $-$0.5$\pm$0.1    & $3.8 \times 10^{-5}$  &    32  & $-$20 &   n.s.   &  n.s.	 &  D  & $-$44.3   \\
WW Vul               &  440     & 3.98    &    1.32  &  1.30   &  2                        & E &  $-$0.07$\pm$0.02  & $1.6 \times 10^{-4}$  &    60  & $-$42 &$|$EW$|$ $<$ 0.05 &  &  D  & $-$24.0   \\
HD 190073           &$>$290$^H$ & 3.95$^F$&    1.72  &  1.92   &  3                        & E &  $-$0.07$\pm$0.02  & $4.7 \times 10^{-4}$  &    25  & $-$12 &$-$0.01  &  13	 &  P  & $-$27.1   \\
LkH$\alpha$ 224      &  980     & 3.89    &    1.86  &  2.32   &  6                        & U & $|$EW$|$ $<$ 0.15  & $< 1.9 \times 10^{-3}$&       &       &$|$EW$|$ $<$ 0.2  & &  P  & $-$17.6   \\
BD$+$46$^\circ$3471  &  1200    & 3.99    &    2.50  &  2.43   &  8                        & E &  $-$0.04$\pm$0.01  & $1.3 \times 10^{-3}$  &    48  & $-$36 &$|$EW$|$ $<$ 0.04 &     &  P  & $-$21.7   \\
SV Cep               &  440     & 3.97    &    1.02  &  1.39   &  4                        & E &  $-$0.12$\pm$0.01  & $1.5 \times 10^{-4}$  &    63  & $-$34 &$-$0.05  &  46	 &inv. P&  $-$10.1   \\
IL Cep               &  690$^H$ & 4.27$^B$&    3.76  &  2.18   &  1                        & U & $|$EW$|$ $<$ 0.02  & $< 2.1 \times 10^{-3}$&       &       &$|$EW$|$ $<$ 0.05 & &  D  & $-$34.5   \\
\hline		       	     		   	      					   						       	                    
\textbf{Group III} & & & & & & & & & & & & & & \\
Z CMa                &  1050    & 4.48    &    5.13  &  3.71   &  36                       & E &  $-$4.3$\pm$0.2    & $3.8$                 &   421  &$-$249 &$-$1.08   &  695	 &  P  & $-$45.2   \\
MWC 297              &  250     & 4.37    &    4.01  &  2.69   &  $-$16                    & E &  $-$6.91$\pm$0.02  & $7.1 \times 10^{-1}$  &    22  &     5 &$-$1.56   &  21	 &  S  &$-$355.3   \\
LkH$\alpha$ 225      &  980     & 3.86    &    2.37  &  3.26   &  6                        & E &  $-$0.3$\pm$0.1    & $1.4 \times 10^{-2}$  &    65  & $-$51 &$|$EW$|$ $<$ 0.2  &  &  U  &         \\
PV Cep               &  440     & 3.91    &    0.76  &  1.92   &  $-$18$^a$                & E &  $-$2.9$\pm$0.1    & $2.7 \times 10^{-3}$  &   321  &$-$164 &$|$EW$|$ $<$ 0.9  &  &  P  & $-$49.9   \\
V645 Cyg             &  3500    & 4.58    &    4.42  &  4.61   &  $-$14                    & E &  $-$5.4$\pm$0.6    & $6.5 \times 10^{-1}$  &   423  &$-$257 &$-$1.44   &  390     &  D  & $-$121.7   \\
\end{tabular}
\normalsize
\end{center}
\end{table*}

\subsection{\OI versus the infrared solid-state bands}

In this section we look for correlations between the PAH parameters
and the \OI 6300\A parameters for the 40 sample objects which were
included in the analysis of AV04.

\begin{figure}
\rotatebox{0}{\resizebox{3.5in}{!}{\includegraphics{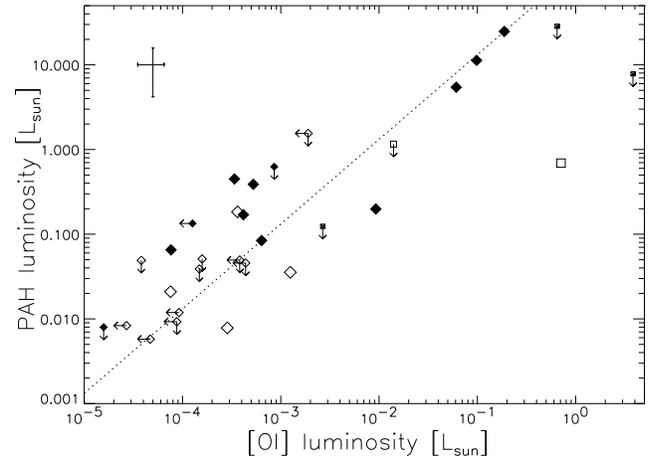}}} 
\caption{ The PAH luminosity versus the \OI 6300\A luminosity. The
  plot includes the 12 group~I, 14 group~II and 5
  group~III sources for which (an upper limit of) both measurements
  was available. The
  dotted line represents the median PAH-over-\OI luminosity ratio
  $L$(PAH)/$L$([O\,{\sc i}]) $= 130$. Plotting symbols are
  as in Fig.~\ref{EWvsFWHM.ps}. The typical error bar is indicated in
  the upper left corner.}
\label{LPAHvsLOI.ps}
\end{figure}

Fig.~\ref{LPAHvsLOI.ps} shows the estimated PAH luminosity versus the
\OI 6300\A luminosity. The 31 objects that are included in this plot
have a pure-emission \OI profile. A clear positive correlation is seen for the
14 stars that have detected PAH as well as \OI 6300\A emission. The
PAH luminosity is on average $350$ times larger than the \OI
luminosity. All group~III sources in this sample have strong \OI
luminosity, but weak or undetected PAH emission. The PAH and \OI
luminosities of the group~II sources are weak or undetected in most
cases, while a typical group~I source displays strong PAH \textit{and}
\OI emission.

\begin{figure}
\rotatebox{0}{\resizebox{3.5in}{!}{\includegraphics{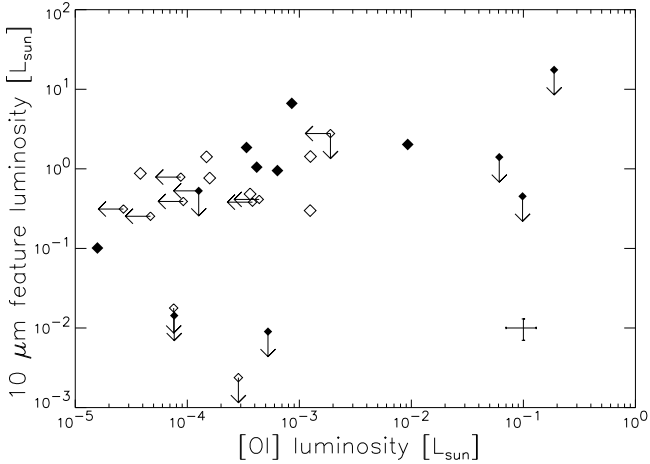}}} 
\caption{ The luminosity of the amorphous 10 micron silicate feature
  versus the \OI 6300\A luminosity. The
  plot includes the 12 group~I and 15 group~II sources for which (an
  upper limit of) both measurements was available. The group~III
  sources are left out since they display silicate \textit{absorption} at
  10$\mu$m. Plotting symbols are
  as in Fig.~\ref{EWvsFWHM.ps}. The typical error bar is indicated
  in the lower right corner.}
\label{LSivsLOI.ps}
\end{figure}

When comparing the luminosity of the amorphous 10 micron silicate
feature to the \OI 6300\A luminosity (Fig.~\ref{LSivsLOI.ps}), no
clear correlation is seen. The arrows which indicate upper limits
support this non-correlation.

%------------------------------------------------------------------

\section{Interpretation \label{secinterpretation}}

A few sources like the group~III outliers \object{Z CMa}, \object{PV
Cep} and \object{V645 Cyg}, but also group~I source \object{HD 200775}
and group~II source \object{HD 141569} have broad features (FWHM
$>$100 \kms) with 
pronounced blue wings. A few more sources have \textit{low}-velocity
blueshifted centroids. Nevertheless, the majority of observed \OI
6300\A emission lines have narrow (FWHM $\sim$ 50 \kms) symmetric
profiles, centered around the stellar radial velocity. Some of the
high-resolution spectra display double-peaked profiles with a
peak-to-peak distance of $\sim$10 \kms. The low velocities,
symmetry of the feature and the peak-to-peak separation correspond to
what one would expect from an emission-line region at the
circumstellar-disk surface. We interpret the observed 
\OI lines in the majority of our sample as being circumstellar-disk emission
features. The forbidden-line emission region is located in the disk's
atmosphere; a warm layer, directly irradiated by the central star and
corotating with the disk. The
blue wing which is present in a minority of the cases is most probably
emanating from an outflow, with the red wing eclipsed by the
circumstellar disk. In this section we will discuss the arguments
for this interpretation.

\begin{figure}
\rotatebox{0}{\resizebox{3.5in}{!}{\includegraphics{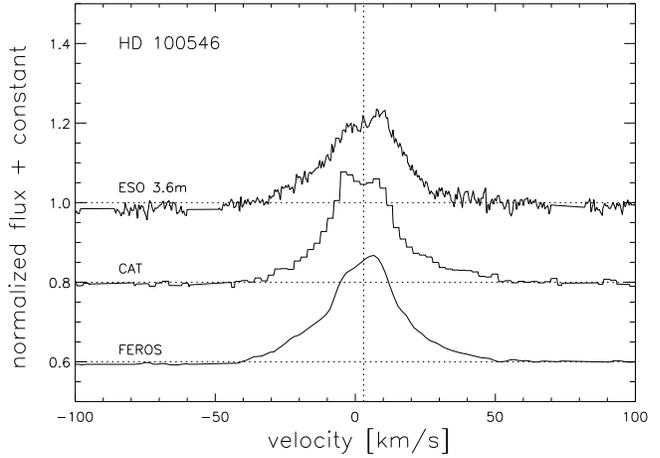}}} 
\caption{ The \OI 6300\A emission line in the spectra of HD
  100546. The double-peaked profile is clearly seen in the ESO 3.6m
  and CAT spectrum. Due to the lower spectral resolution, the profile
  remains unresolved in the FEROS spectrum. The horizontal dotted
  lines represent the continuum levels of the three spectra. The
  vertical line is the mean centroid position of the feature at 3 \kms.}
\label{HD100546.ps}
\end{figure}

When investigating the narrow profiles of the non-blueshifted
features in the high-resolution ESO 3.6m spectra, a double-peaked
profile is seen in a some cases. In Fig.~\ref{HD100546.ps} this
specific shape in the spectrum of \object{HD 100546} (group~I) is
shown. The two peaks are located at equal distance ($\sim \pm$6 \kms)
from the centroid position. Similar profiles have been observed in CO
emission lines \citep[e.g.][]{chandler, thi} and have been attributed to the
Keplerian rotation of the circumstellar gas disk. Also in group~I
sources \object{HD 97048}, \object{HD 135344} and \object{T CrA}, for
which we have ESO spectra, this double-peaked shape is clearly
seen. The other spectra have a lower spectral resolution, and thus the
double peaks are less clear. Considering the latter, more sources with
symmetrical low-velocity profiles, like \object{AB Aur}, might have a
double-peaked line shape. In section~\ref{secmodeling} we model the
spectral profiles of some of these features.

A second indication that the forbidden-line emission region is
connected to the circumstellar disk lies in the correlation between the
strength of the \OI line and the SED classification of the sources. 
As explained before, group~I sources are suggested to have a
flared-disk geometry, while it is believed that group~II members have
self-shadowed disks.
The disk geometry can play an important role in the amount of neutral
oxygen in the upper level of the 6300\A line. Apart from thermal excitation,
photodissociation of OH and H$_2$O molecules by UV photons can
significantly increase the number of excited oxygen atoms
\citep{vandishoeck84}. When these molecules are exposed to high UV
fluxes, e.g. at the surface of a flared disk,
the non-thermal population of the upper level of the 6300\A line
occurs very efficiently. The outer parts of self-shadowed disks lie
completely in the shade of the puffed-up inner rim. Consequently, the
UV radiation of the central star cannot reach the disk surface and no
significant photodissociation of OH and H$_2$O molecules will take
place. This would result in weaker \OI emission for group~II sources.
The absence of the feature in 40\% of the group~II sources and
Fig.~\ref{histogramEW.ps} indeed show
that the correlation between \OI 6300\A luminosity and M01
classification exists, as described in \textsection\ref{oivssed}.

Furthermore, the \OI luminosity is also correlated to the PAH
luminosity of the source (see Fig.~\ref{LPAHvsLOI.ps}). These
molecules need stellar UV photons to get excited. In flared disks,
significantly more PAH emission is observed 
than in self-shadowed disks (AV04). The observed correlation between
the \OI and PAH luminosity suggests that both the oxygen atoms and the
PAH molecules radiate from the same location: the disk's atmosphere.
This low-density, UV-photon-immersed region at the disk's
surface offers the ideal locus for the photodissociation of OH and
H$_2$O and the consequent 6300\A forbidden-line emission.

The non-correlation of the \OI 6300\A luminosity with the luminosity
of the amorphous 10 micron band is also expected under the present
hypothesis. The small silicate grains that cause this emission feature
are thermally excited and do not need direct stellar flux. The
observed lack of correlation is similar to the non-correlation between
the PAH luminosity and the luminosity of the 10 micron feature as
described by AV04.

%------------------------------------------------------------------

\section{Modeling the \OI emission region \label{secmodeling}}

In this section we model the \OI emission in the atmosphere of a
flared disk. We have computed the structure of a flared disk
\citep{chiang} and determined the layer from which the \OI emission
emanates. In this layer, the density is 
low and the UV flux abundant. We calculate the intensity of the \OI
emission for different stellar and disk parameters. In order to model
the line profiles observed in the spectra, the theoretical
emission-line profile was determined by convolving the computed
intensities with a Keplerian-rotation profile.

\subsection{The flared-disk model}

We have implemented the flared-disk model of \citet{chiang} with some
improvements described by \citet[][their section 2.1.1 and 2.1.2]{dullemond01}. The
input values of the model consist of the stellar parameters
($M_\star$, $T_\star$ and $L_\star$), dust opacities $\kappa_\nu$ and
the surface density $\Sigma$. The latter quantity is
assumed to be a power-law function of the radius $R$ to the star:
$\Sigma=\Sigma_0 \left( R [\AU] \right)^{-\beta}$. The surface
density at 1 AU ($\Sigma_0$) and the power ($\beta$) can be
chosen freely. The structure of the disk's interior is then calculated
iteratively by demanding vertical hydrostatic equilibrium at each
radius. The output quantities \citep[in the nomenclature
  of][]{dullemond01} include the pressure scale height $h_{cg}$, the
disk surface height $H_{cg}$, the midplane temperature $T_i$ and the surface 
temperature $T_s$. These are all a function of the radius $R$. The
density $\rho$ is a Gaussian 
function in the vertical (i.e. $z$) direction, centered around the midplane
($z=0$). Its half-width depends on the pressure scale height
$h_{cg}$, which is a function of the midplane temperature $T_i$.

\begin{figure}
%\rotatebox{0}{\resizebox{3.5in}{!}{\includegraphics{2484fig18color.ps}}} 
\rotatebox{0}{\resizebox{3.5in}{!}{\includegraphics{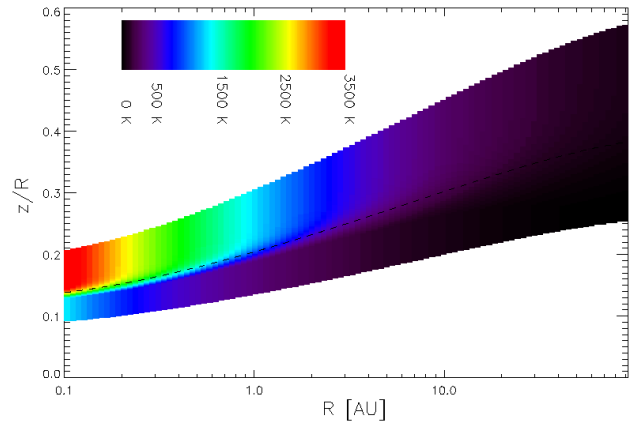}}} 
\caption{ The temperature structure of a flared-disk model. On the
  x-axis the radial distance $R$ to the star is plotted, on the y-axis the
  ratio of the height above the midplane $z$ over $R$. The color
  scale is linear and ranges from 0 to 3500K. The dashed line
  represents the disk surface height $H_{cg}$.}
\label{tempTEMP.ps}
\end{figure}

The temperature distribution in the disk is calculated more precise
than described in \citet{chiang} and \citet{dullemond01}. These authors assume a
two-temperature model which is determined by the midplane temperature $T_i$
and the temperature at the optically thin surface layer
$T_s$. Inspired by the full-fletched, computationally demanding models
of \citet{dullemond02} and \citet{dullemond04}, we allowed for a
temperature gradient 
in vertical direction. The stellar flux penetrates the disk, directly
heating the circumstellar matter. At a certain radius $R$ from the
star and height $z$ above the midplane, the temperature $T(R,z)$ can be
determined using 
\begin{eqnarray} 
\lefteqn{\int_0^\infty B_\nu (T(R,z))\, \kappa_\nu\,  d\nu  =} & &
     \nonumber \\ 
& & \qquad  \qquad \frac{R_\star^2}{4 R^2} \int_0^\infty B_\nu (T_\star)\, 
\exp \left( -\tau_{\mathrm{rad}}(R,z) \right)\,
\kappa_\nu\,  d\nu 
\label{disktemp} 
\end{eqnarray}
in which $R_\star$ and $T_\star$ are the stellar radius and
temperature. The radial optical depth $\tau_{\mathrm{rad}}$ can be
approximated by the ratio of the vertical optical depth
$\tau_{\mathrm{vert}}$ and the flaring angle of the disk $\alpha$:
\begin{equation}
\tau_{\mathrm{rad}}(R,z) = \frac{\tau_{\mathrm{vert}}(R,z)}{\alpha(R)}. 
\end{equation}
This approach is similar to the one used to compute the surface scale
height of the disk in the original \citeauthor{chiang} 
model. At the surface, where $\tau_{\mathrm{rad}} \ll 1$, the
temperature $T(R,z)$ is equal to the optically thin $T_s(R)$. Deeper in the disk's
interior, the $T(R,z)$ computed from equation (\ref{disktemp}) drops
until it reaches the midplane temperature $T_i(R)$. At these
locations, the dominant 
heating source is not the \textit{direct} stellar flux, but the
flux radiated downward from the disk surface. The latter indirect flux
maintains the midplane temperature $T_i$ and is only important in
regions were no direct stellar flux can penetrate. Therefore, we have
set all $T(R,z)$ equal to $T_i(R)$ at these locations. In
Fig.~\ref{tempTEMP.ps} the temperature structure of a typical
flared-disk model is shown.

We note that the temperature stratification in the
disk and the density function are not computed self-consistently.
The temperature gradient is calculated \textit{a posteriori};
the effect of the higher temperatures in the upper layers of the
disk is not included in the determination of $\rho$. The deviation
from the self-consistent calculation is however very small.

\subsection{The \OI emission}

To determine the \OI 6300\A emission emanating from a flared circumstellar
disk, we have implemented the model of \citet{storzer00} for \OI
emission. They have modelled the optical forbidden-line emission from
a plane-parallel semi-infinite photodissociation region (PDR). Both
thermal and non-thermal emission are included. Slight 
changes were applied to the method to adapt it to our specific case.

\subsubsection{Thermal \OI emission}

Following \citet{storzer00}, a five-level oxygen atom is assumed. The
atomic data and references can be found in their paper
(Table 1 of the Appendix). Since we are only interested in modeling the 6300\A
line, the relevant transitions for the present paper are ${^3P_1}
\longrightarrow {^3P_2}$ (63.2~$\mu$m), ${^3P_0} \longrightarrow
{^3P_2}$ (44.2~$\mu$m), ${^1D_2} \longrightarrow {^3P_2}$
(6300.3\r{A}) and ${^1S_0} \longrightarrow {^1D_2}$
(5577.4\r{A}). Collisions with free electrons and atomic hydrogen are
considered. From PDR models \citep[e.g.][and references 
  therein]{hollenbach99}, one derives that the transition region from
the atomic-H dominated upper layers to the disk interior where H$_2$
is abundant occurs at about $A_V = 1$. Based on the latter, we assume
that only at $A_V<1$ free electrons and H atoms are important
collisional partners for the O {\sc i} atom. In optically thick
regions, the collisional rate is set to zero.

The proton number density $n_p$ depends on the radial and vertical
position $(R,z)$ in the disk and the input surface density $\Sigma$. It
is computed from the flared-disk model. The oxygen number density is
the product of the proton density and the fractional oxygen abundance
\begin{equation}
n(O) = n_p \left[\frac{O}{H}\right].  
\end{equation}
A typical interstellar value for $[O/H]$ is $5\times10^{-5}$
\citep[e.g.,][]{allen76}. 
In the region where oxygen is not ionized, the atomic hydrogen is neutral
as well, because the ionization energy of the two species is
comparable. In this region the free electrons are almost all due to the
ionization of neutral carbon atoms. Assuming all carbon atoms in the
considered region are singly ionized, the free-electron number density is 
\begin{equation}
n_{e^-} = n_{C^+} = n_p \left[ \frac{C}{H} \right].
\end{equation}
A typical interstellar fractional abundance of carbon is $3\times10^{-4}$.

To determine the thermal \OI 6300\A emission, the population of the upper
level of this line ($^1D_2$) needs to be known. Using the formula of
thermal equilibrium (see Appendix), one can determine the relative
population of two levels $a$ and $b$ for a temperature
$T$. From the latter ratios, 
the relative fraction $F$ of O {\sc i} atoms in the $^1D_2$ state can be
calculated. The number density of the thermally populated upper level
of the 6300\A line is $n_{u, th} = F\, n(O)$.

Since the typical temperature range (10--1500K) in our disk models is
rather low to thermally excite the oxygen atoms, the emanating \OI 6300\A
emission is very weak. The intensity of the thermal emission is more
than a million times smaller than the non-thermal \OI emission, which is
discussed in the next section.

\subsubsection{Non-thermal \OI emission \label{secnonthermal}}

Following \citet{storzer00}, the dominant non-thermal excitation
mechanism for neutral oxygen atoms is the photodissociation of OH
molecules. This results in a hydrogen atom and an excited oxygen
atom. A fraction of the latter \citep[$\sim$55\%,][]{vandishoeck84}
find themselves in the upper state ($^1D_2$) of the 6300\A line.
This mechanism hence produces strong non-thermal \OI emission in
regions where the photodissociating UV flux is abundant and the
densities high enough to have a sufficient amount of emitting oxygen
atoms.

In our simple model for non-thermal emission, we assume that the
fractional OH abundance $\epsilon(OH)$ is constant throughout the
disk. In reality, this may not be the case, as the optically thick disk
interior will have a much higher $\epsilon(OH)$ than the
photon-immersed surface layers. Nevertheless this assumption needs not
to be valid for the entire disk, but only for the \OI emission
region. The latter will be located close to the $\tau=1$ surface and
is geometrically quite thin, because the proton density drops off rapidly
when moving away from the disk midplane while the high optical depth
in the disk interior prohibits photodissociation of the OH
molecules. From the OH number density $n(OH) = \epsilon(OH) \times
n_p$, the density of non-thermally-excited oxygen 
atoms can be determined \citep[][adapted to the UMIST rate
coefficients]{storzer98}: 
\begin{equation}
 n_{u, non-th} = f\, G_0\, \frac{3.50 \times 10^{-10}}{A}\, e^{-1.7
    A_V}\, n(OH).  \label{nunonth}
\end{equation}7
In this formula, $f$ ($\approx$ 55\%) is the fraction of oxygen atoms
released by the photodissociation of OH in the upper level of the
6300\A line. The sum ($A$) of the Einstein transition rates from the
${^1D_2}$-level downwards is approximately $8.4 \times 10^{-3}$ s$^{-1}$
\citep{osterbrock89}.

\begin{figure}
%\rotatebox{0}{\resizebox{3.5in}{!}{\includegraphics{2484fig19color.ps}}} 
\rotatebox{0}{\resizebox{3.5in}{!}{\includegraphics{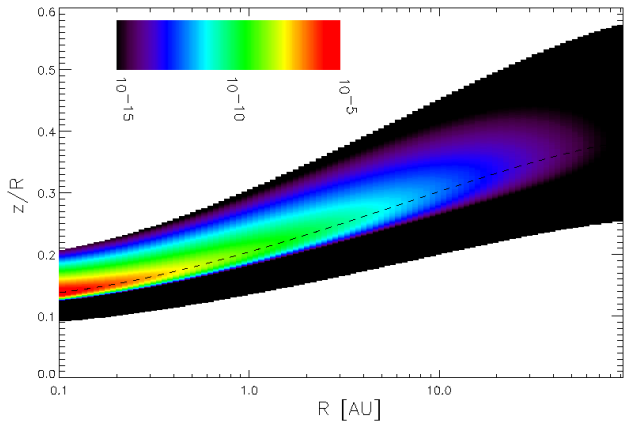}}} 
\caption{ The \OI 6300\A emissivity in the flared-disk
  model. The x-axis gives the radial distance $R$ to the central star in
  AU, the y-axis plots $z/R$, the ratio of the vertical height and the
  radial distance. The colors represent the local \OI 6300\A emissivity
  on a log-scale which covers 10 orders of magnitude. On the color 
  bar, the emissivity is given in [erg s$^{-1}$ cm$^{-3}$
  sr$^{-1}$]. The disk region sampled by the model 
  ranges from roughly 0.6 to 1.5 times the disk surface height
  $H_{cg}$. The latter is indicated by the dashed line. The
  geometrical full width at half maximum of the \OI emission region in
  the vertical direction is 5--10\% of $H_{cg}$.}
\label{tempOI.ps}
\end{figure}

The local \OI 6300\A emissivity $j(R,z)$ (in erg
s$^{-1}$ cm$^{-3}$ sr$^{-1}$) is given by
\begin{equation}
 j(R,z) = \frac{1}{4 \pi} h\nu\ A_{6300} n_u(R,z)
\end{equation}
The Einstein coefficient $A_{6300}$ is $6.3 \times 10^{-3}$
s$^{-1}$ \citep{osterbrock89}. The value $h\nu$ is the energy of a 6300\A photon.
The density $n_u$ is the sum of the densities of the thermally and non-thermally
excited oxygen atoms. In Fig.~\ref{tempOI.ps}, $j(R,z)$ is plotted as
a function of the vertical and radial location in the disk. The dashed
line represents the disk surface height $H_{cg}$ of the disk. The
emission is located around this region.

Integrating the emissivity over the vertical direction, considering the
optical depth $\tau_{6300}$ at 6300\r{A}, one extracts the surface intensity $I(R)$
at each radius $R$:
\begin{equation}  
I(R) = \int_z j(R,z)\, e^{-\tau_{6300}}dz  
\end{equation}
The unit of $I(R)$ is erg s$^{-1}$ cm$^{-2}$ sr$^{-1}$. Note that, due to the
shallow angle under which the stellar light impinges on the disk, one
looks much deeper into the disk in the \textit{vertical} than in the
\textit{radial} direction. When looking at the
disk face-on, the observer practically sees the entire \OI emitting
region (on one side of the disk).

\subsection{The line profiles}

\begin{figure}
\rotatebox{0}{\resizebox{3.5in}{!}{\includegraphics{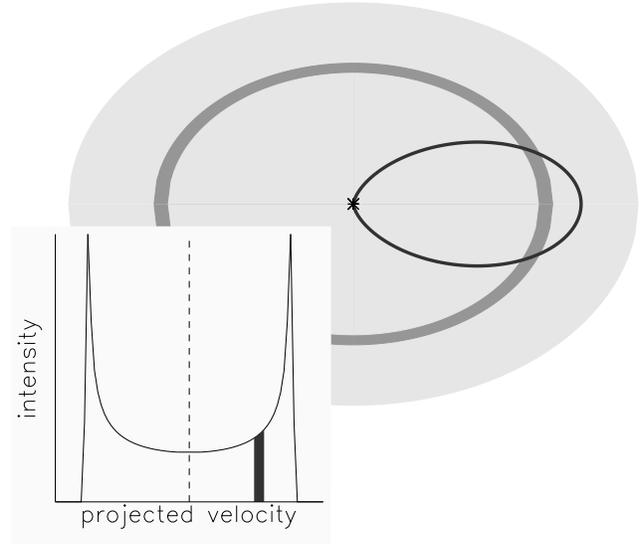}}} 
\caption{ Pictogram of a flat rotating circumstellar disk, seen under
  an inclination of $i=45^\circ$. The location of the central star is
  indicated by the asterisk. The light-grey band represents a
  circular orbit. In this orbit, the rotational velocity as well as
  the \OI intensity (I$_o$) are constant due to the assumption of
  axisymmetry. The narrow dark-grey band represents the region where
  the \textit{projected} rotational velocity is equal to a certain value
  $v_p$. At the cross-section of both surfaces the projected velocity is
  equal to $v_p$ and the intensity is equal to I$_o$. The insertion shows
  the intensity-versus-velocity profile of the light-grey band. The
  dark-grey part represents the contribution of the cross section of
  the dark-grey band and the orbit in the disk.}
\label{pictogram.ps}
\end{figure}

To convert the intensity-versus-radius function $I(R)$ to an
intensity-versus-velocity profile, we assume that the rotation
of the disk is Keplerian\footnote{The IDL code \textit{keprot.pro},
which converts an intensity-versus-radius profile into an
intensity-versus-velocity profile, can be downloaded at \\
\textbf{http://www.ster.kuleuven.ac.be/$\sim$bram/OI/keprot.pro}
\textbf{http://www.ster.kuleuven.ac.be/$\sim$bram/OI/keprot.README}}.
We anticipate this to be a fairly accurate assumption, since the disk mass
is expected to be much smaller than the mass of the central star
\citep[e.g.,][]{ackesubmm}. 
Additionally we adopt that the Keplerian rotation velocity of matter in
the upper layers of a flared disk is the same as the rotational
velocity in the midplane. 
Fig.~\ref{pictogram.ps} illustrates how the profile is 
computed. The light-grey band in the pictogram
represents a circular orbit in the inclined --geometrically flat--
disk where the \OI intensity is constant. The 
disk's matter in this band rotates around the central star with a
Keplerian velocity $v_{\kep} = \sqrt{GM_\star /
  R_{\mathrm{orbit}}}$ in which $R_{\mathrm{orbit}}$ is the radius of
the orbit and $M_\star$ is the mass of the central star. The 
narrow darker-grey band represents the parts of the disk that move
towards (or away from) the observer with the same \textit{projected} velocity
$v_{\proj}$. This area is the region where
\begin{equation}
 R = \frac{GM_\star}{v_{\proj}} \sin^2 i\, \sin^2 \theta 
\end{equation}
with $\theta$ the angle in the disk plane ($\theta=0^\circ$
towards the observer). The inclination $i$ is defined to
be $0^\circ$ for pole-on disks. The black
surface is the cross-section of both bands. By adding up these
regions and multiplying with the intensity, the
intensity-versus-velocity profile is built up for each projected
(=observed) velocity.

The computed theoretical profile is then convolved with a Gaussian
function\footnote{The IDL code \textit{convolve.pro}, which convolves
an intensity-versus-velocity profile with a Gaussian function, can be
downloaded at \\
\textbf{http://www.ster.kuleuven.ac.be/$\sim$bram/OI/convolve.pro}
\textbf{http://www.ster.kuleuven.ac.be/$\sim$bram/OI/convolve.README}}.
This function mimics the effect of instrumental 
broadening. The width of the --unresolved-- telluric lines in each
type of spectrum were used as the width of the instrumental Gaussian
profile. 

\subsection{Discussion of the model parameters}

The input parameters for our model that calculates the non-thermal
emission consist of some 
stellar parameters ($L_\star, L_\uv, M_\star, T_\star$) which are
derived from the photometric data. Furthermore, a dust opacity table
and dust-to-gas mass ratio are needed. Following the
\citet{dullemond01} model for \object{AB Aur}, we assume
that the dust consists of olivines and represents 1\% of the total
disk mass. The stellar parameters are $L_\star = 47~L_\odot$,
$L_\uv = 36~L_\odot$, $M_\star = 2.4~M_\odot$ and $T_\star = 9520$K. 
The free parameters in our \OI emission model are the surface density
$\Sigma = \Sigma_0\, (R[AU])^{-\beta}$, the fractional OH abundance
$\epsilon(OH)$ and the disk's inclination $i$. In the template model,
we take $\Sigma_0 = 10^4$ g cm$^{-2}$, $\beta = 2$, $\epsilon(OH) = 1
\times 10^{-6}$ and $i = 45^\circ$. The inner and outer radius of the
disk, $R_{in}$ and $R_{out}$, are 0.1 AU and 100 AU respectively.
We discuss the influence of the four free input parameters on the
shape of the emission profile, starting with the surface density. 

\begin{figure}
\rotatebox{0}{\resizebox{3.5in}{!}{\includegraphics{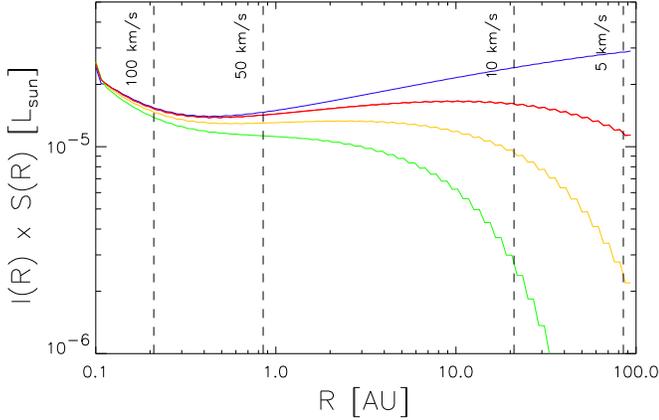}}} 
\caption{ The intensity-versus-radius distribution $I(R)$ for four
  template models, with different values for the surface-density
  parameters. From top to bottom, $\beta$ is 1.0, 2.0, 2.5 and 3.0
  respectively. The value of $\Sigma_0$ is chosen such that the disk
  masses of the four models are the same. The intensity function is
  multiplied by the total surface 
  $S(R)$ of the ring at radius $R$. In this way the fluxes from the
  inner and outer parts of the disk can be compared directly. The vertical dashed
  lines mark the radii where the Keplerian velocity is 5, 10, 50 and
  100 \kms. The \textit{shoulder} of the intensity
  distribution shifts closer to the star with increasing $\beta$. Furthermore,
  the \textit{total} emission is smaller when the power-law describing $\Sigma$
  has a steeper slope.}  
\label{sigbet.ps}
\end{figure}

\begin{figure}
\rotatebox{0}{\resizebox{3.5in}{!}{\includegraphics{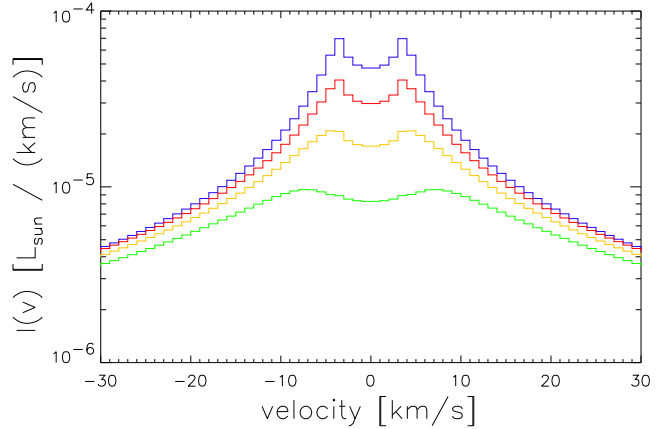}}} 
\caption{ The line profiles $I(v)$ corresponding to the
  intensity-versus-radius distribution in Fig.~\ref{sigbet.ps}. From
  top to bottom, $\beta$ is 1.0, 2.0, 2.5 and 3.0
  respectively. Since the intensity-versus-radius distribution falls
  off rapidly in the models with large $\beta$ values, the
  corresponding profiles have peaks at higher velocities than the
  profiles belonging to low-$\beta$ models.}
\label{sigbetprof.ps}
\end{figure}

The two parameters $\Sigma_0$ and $\beta$ describing the power-law
surface density are coupled when the disk mass is known: 
\begin{equation} 
 M_{disk} = 2 \pi \int_{R_{in}}^{R_{out}} \Sigma_0\, R^{1-\beta} dR 
\end{equation}
The disk mass of our template model is $M_{disk} = 0.049
M_\odot$. In Fig.~\ref{sigbet.ps}, the intensity-versus-radius
distribution of the \OI 6300\A emission is plotted for four
models. The input parameters of the latter are the 
template values, except for $\Sigma$. The power $\beta$ ranges from
1.0 to 3.0, while $\Sigma_0$ is appropriately adapted to ensure that
the disk mass remains unaltered. The total emission intensity
decreases with increasing $\beta$. Fig.~\ref{sigbetprof.ps} shows the
line profiles, corresponding to the intensity distributions in
Fig.~\ref{sigbet.ps}. 

The fractional OH abundance $\epsilon(OH)$ scales linearly with the
total intensity of the profile. It does not alter the 
shape, as we assume it is constant throughout the \OI emission
region. 

\begin{figure}
\rotatebox{0}{\resizebox{3.5in}{!}{\includegraphics{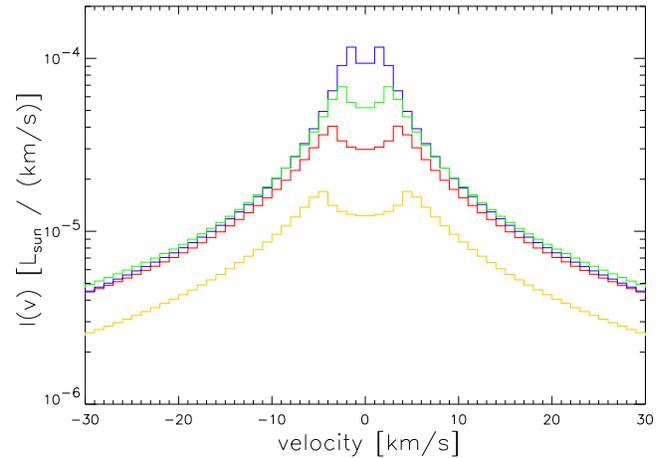}}} 
\caption{ The line profiles $I(v)$ of the template model seen under
  different inclinations. From top to bottom, the inclination is
  $18^\circ, 30^\circ, 45^\circ$ and $70^\circ$ respectively. Note the decrease
  in integrated intensity and the shift of 
  the peaks with increasing $i$.}
\label{inclprof.ps}
\end{figure}

Observations of Doppler-broadened spectral lines do not observe the
real, but the projected velocities in the system. In the present
flared-disk model, the inclination $i$ is a free parameter which alters
the shape as well as the intensity of the computed profile. The total
integrated intensity $\int I(v) dv$ is proportional to $\cos i$, while
the position of the two typical peaks of the line changes with $\sin
i$. Fig.~\ref{inclprof.ps} shows the line profiles of the template model as seen
under different inclinations. Note that inclinations larger than
about $70^\circ$ are not relevant in flared disks, since at such
high $i$, the outer parts of the disk would occult most of the disk
surface.

%------------------------------------------------------------------

\section{Comparison with the observations \label{seccompare}}

In this section we compare our model results to the observations. We
focus on the sample stars which display a narrow, symmetric and
centered profile. Since the thermal emission is weak compared to the
non-thermal emission for the relatively cool disks studied in this
paper, we will only consider the non-thermal emission component in the
following discussion. 

\begin{figure}
\rotatebox{0}{\resizebox{3.5in}{!}{\includegraphics{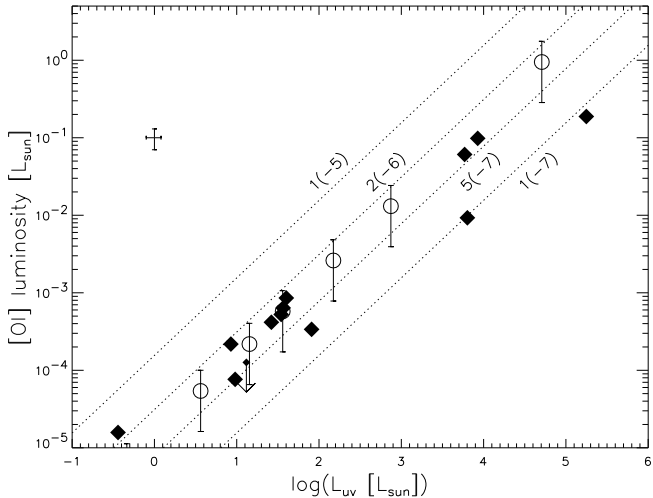}}} 
\caption{ Similar plot as Fig.~\ref{LUVvsLOI.ps}. The filled diamonds
  are the observed \OI intensities for the group~I sources. The open
  circles represent model results. The typical error bar for the
  observations is given in the upper left corner. The models differ
  from each other 
  in stellar parameters; from the lower left to the upper right, the
  model [O\,{\sc i}]-versus-UV luminosity for a
  typical F0V, A3V, A0V, B8V, B5V and B0V star are plotted
  respectively. The error bars on the model determinations represent
  the spread in results when simultaneously varying the surface
  density $\Sigma$ and 
  inclination $i$ in the ranges described in Figs.~\ref{sigbetprof.ps}
  and \ref{inclprof.ps}. The dotted lines represent the model results
  when altering the fractional OH abundance $\epsilon(OH)$ and using
  the average [O\,{\sc i}]-versus-UV luminosity ratio of the
  models. The abundance is indicated next to each line, with $a(-b)$
  representing $a \times 10^{-b}$. The fractional OH abundance of the
  models is $1 \times 10^{-6}$.} 
\label{LUVvsLOImodel.ps}
\end{figure}

\begin{figure}
\rotatebox{0}{\resizebox{3.5in}{!}{\includegraphics{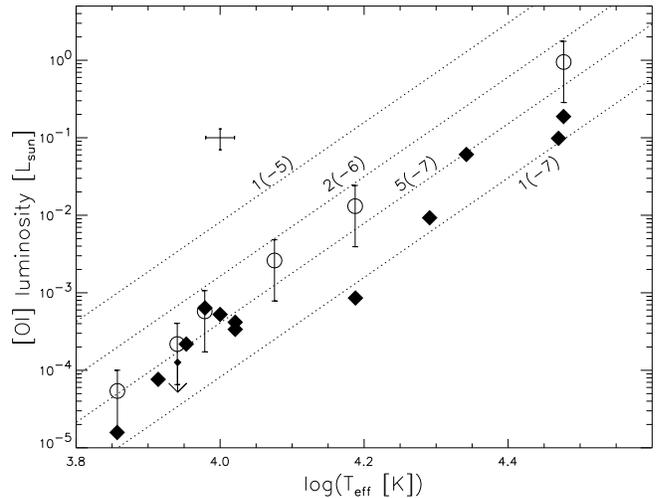}}} 
\caption{ The \OI 6300\A luminosity versus the effective temperature
  of the group~I sources (filled diamonds). The open circles and
  dotted lines plotted in this figure represent the same models as in
  Fig.~\ref{LUVvsLOImodel.ps}. }
\label{TeffvsLOImodel.ps}
\end{figure}

We compare the observed \OI 6300\A emission luminosity and its
theoretical counterpart to the UV luminosity in
Fig.~\ref{LUVvsLOImodel.ps}. Our model 
applies for flared-disk geometries, therefore only the group~I sources
have been plotted in the figure. The model results and the
observations agree nicely for an acceptable range of the free
parameters $\Sigma$, $i$ and $\epsilon(OH)$. Note that all models in
the figure have the same disk mass as the template model. Varying the
disk mass affects the \OI 6300\A emission in a comparable manner as
when the surface density is altered. Fig.~\ref{TeffvsLOImodel.ps}
compares the \OI luminosity to the effective temperature of the group
I objects and the models. Again, reasonable values for the free
parameters lead to theoretical \OI luminosities close to the observed
values.

\begin{figure*}
\rotatebox{0}{\resizebox{\textwidth}{!}{\includegraphics{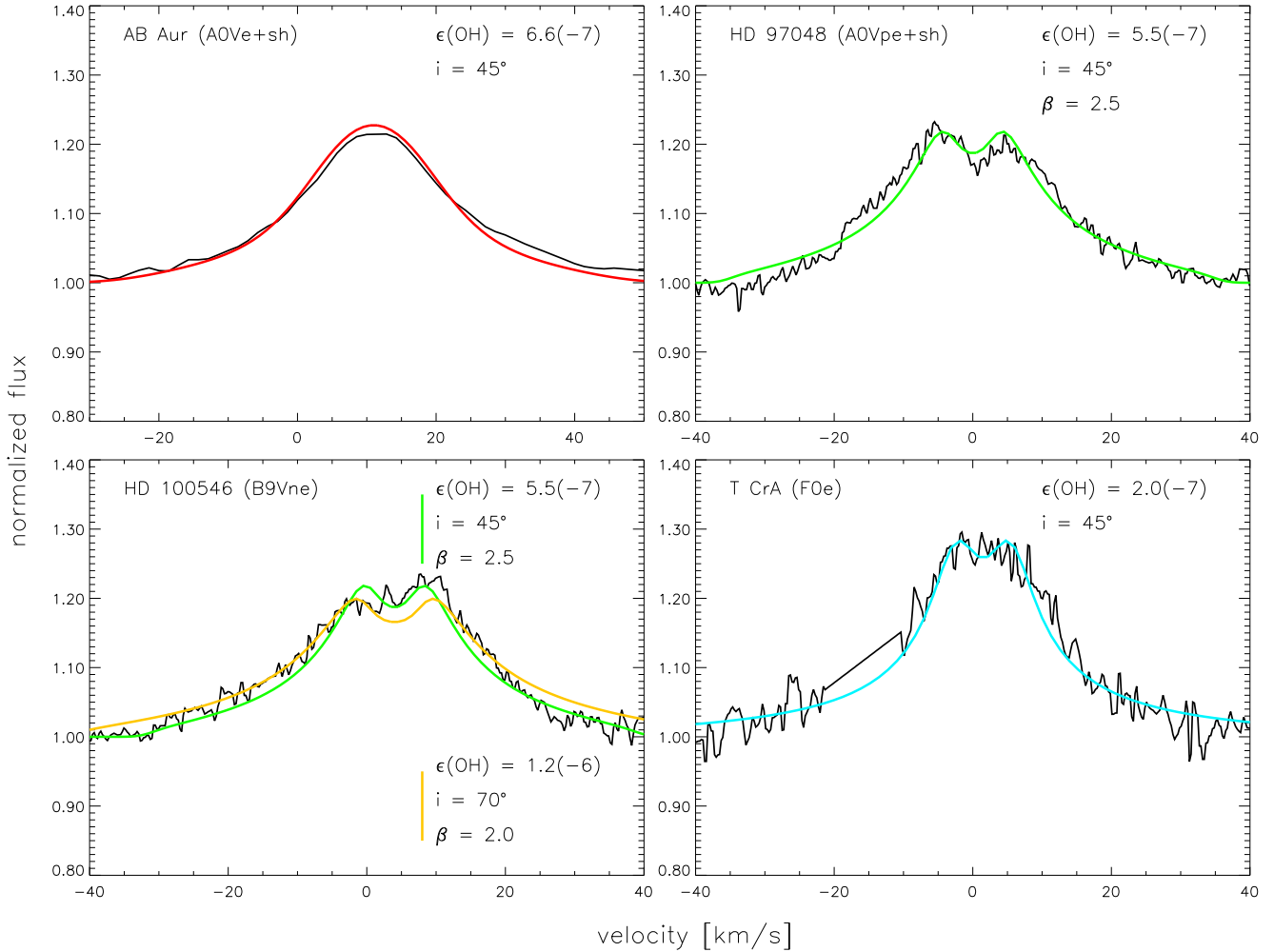}}} 
\caption{ The observed spectra of group~I sources \object{AB Aur},
  \object{HD 97048}, \object{HD 100546} and \object{T CrA} overplotted
  with the model results. The parameters of the model are adapted to
  the stellar parameters of each source combined with the template
  parameters, unless otherwise indicated in the upper right
  corner. We have shifted the theoretical profile to the centroid
  position of the 
  observed feature. For \object{HD 100546} two models with different parameter
  sets have been displayed to illustrate the difficulty to distinguish
  between models.
  The theoretical profiles extracted from our flared-disk model fit
  the observed features strikingly well for a range of reasonable
  model parameters. $a(-b)$ represents $a \times 10^{-b}$.} 
\label{specmod.ps}
\end{figure*}

Not only the strength of the \OI 6300\A emission can be reproduced
with our model. After scaling the total theoretical to the observed
intensity, the typical FWHM and double-peaked shape of the narrow centered
features can be explained by assuming Keplerian rotation of
the flared disk. In Fig.~\ref{specmod.ps} we have overplotted observed
profiles with theoretical ones. For each star, we have used the
stellar parameters ($L_\star, T_\star, M_\star$) as input
parameters. The remaining parameters were set to the template values,
unless otherwise indicated in the plots. Note that the theoretical
profiles are not \textit{fitted} to the data, but are just overplotted
model profiles, rebinned to the spectral resolution of the observed
spectrum. The similarity between theory and observation is
nevertheless striking. For \object{AB Aur} and \object{HD 100546}, the
inclination of the disk have been determined to be $i=76^\circ$
\citep{mannings97} and $i=51^\circ$ \citep{augereau01}
respectively. The observed profiles of the \OI 6300\A line can be
approximated by theoretical line profiles for these inclinations when
using $\epsilon(OH) \approx 2 \times 10^{-6}$ and $\beta = 1.5$ for
\object{AB Aur} and $\epsilon(OH) \approx 6 \times 10^{-7}$ and $\beta
= 2.5$ for \object{HD 100546}. 

The flaring of the disk is due to the dust opacities, while the gas
opacities are negligible. Since dust evaporates above the dust
sublimation temperature $T_{subl} \sim 1500$K, we
have shifted the inner edge of the flared-disk from $R_{in} = 0.1$ AU
to the radius where the surface temperature $T_s$ equals
$T_{subl}$. For a typical F0V, A0V and B0V star, the 
inner radius hence becomes 0.2, 0.8 and 30 AU respectively. As a
consequence, the high-velocity ($|v| \gtrsim$ 50 \kms) wings of the
theoretical \OI emission line profile disappear, in agreement with the
observed spectra of narrow 6300\A features which do not display
extended wings (e.g., \object{AB Aur}, \object{HD 97048}, \object{HD
  100546}, \object{HD 135344}, \object{HD 169142}, \object{R CrA},
\object{T CrA}, \object{HD 179218}, \object{HD 190073}). Furthermore,
the models of \citet{dullemond02} suggest 
that the inner rim of the disk is puffed-up and casts its shadow of
the first few AU in a flared geometry. This effect could be mimicked
in our flared-disk model by shifting the inner radius $R_{in}$ even
further out (up to $\sim$5 AU). The effect of the latter on the
theoretical line profile would be mostly a decrease of the
wings, without reducing the total intensity much.

The innermost parts of the circumstellar disk, where no dust can survive
due to the high temperatures, are relatively small in spatial
dimensions. Nevertheless, we cannot exclude that the contribution of these
parts to the total \OI emission strength is significant. Specifically,
a few group~II sources (e.g., \object{VX Cas}, \object{V586 Ori},
\object{HD 95881}, \object{HD 98922}, \object{HD 101412}, \object{WW
Vul}, \object{SV Cep}) display broader, but relatively
weaker \OI profiles. In our current interpretation and with no external
bright UV source near, this \OI emission cannot emanate from the disk
surface for the self-shadowing prohibits direct stellar flux to reach
this area. In the latter cases, the emission might come from a
rotating gaseous disk \textit{inside} the dust-sublimation radius.
The modeling of this inner gaseous region is however beyond the scope
of the present paper. We refer to the recent paper of
\citet{muzerolle04} for a more elaborate discussion on this subject.

%------------------------------------------------------------------

\section{Conclusions and discussion \label{secconclusions}}

The observed \OI 6300\A emission line in many HAEBE stars in our
sample shows evidence for a \textit{rotating} forbidden-line emission
region. In this
paper we suggest that the surface of the flared circumstellar disk
around the group~I objects is the perfect location to harbor this
emission region. The combination of the direct stellar UV flux and the
relatively high densities in this region give rise to strong 
non-thermal \OI emission, which can explain the observed luminosities
reasonably well. Furthermore, the shape of the spectrally resolved 6300\A
profile in the observations and the profile produced by our
simple model indicate that Keplerian rotation indeed is the broadening
mechanism for (at least the narrow component of) this line. The
observed 6300\A spectra of the group~I members in our sample can be
reproduced strikingly well.

In the group of self-shadowed-disk sources, significantly more targets
do not display the \OI 6300\A emission line (43\% versus 8\% in
group~I). The line profiles of the \textit{detected} \OI 6300\A emission
feature in group~II are twice as weak as, and somewhat broader than the
lines observed in 
group~I spectra. In our current interpretation, the surface of
a self-shadowed disk's outer parts is not directly irradiated by the
central star. However, the non-thermal \OI line formation mechanism
|photodissociation of OH and H$_2$O| may produce the observed group~II
emission lines as well. Assuming that the \OI emission emanates from
the inner gaseous disk naturally explaines the weaker emission and
the higher-velocity wings of the feature: the inner gaseous disk
provides a smaller emission volume and is located closer to the
central star, where the Keplerian velocities are larger.

In T~Tauri stars, the less massive counterparts of HAEBE stars, the
observed \OI emission profile can be explained using the model of
\citet{kwan} \citep[e.g.][]{hartigan95}. However, this model assumes
that a ``super-heated'' disk atmosphere, fed by accretion, is present,
in which the temperatures are sufficiently high ($\sim$10$^4$K) to
produce thermal \OI 
emission. The absence of a strong UV excess and strong photospheric
veiling, the relatively weak IR recombination lines and CO emission,
and the presence of 10 micron silicate emission (AV04) in the majority
of the group~I and II sources in our sample all indicate that these
disks are passive. Furthermore, the observed upper limits for the 
5577/6300 \OI ratio indicate that the observed emission
cannot be thermal. Each model which models the \OI emission based on
thermal processes |including the disk wind model| will have
difficulties reproducing the strength of the \OI 6300\A line in HAEBE
stars. The non-detections of the [\SII] 6731\A line in the spectra of
group~I and II sources in our sample confirm this picture. Moreover,
amongst the late-type 
sample sources no strong \OI emitters are present. In the T Tauri
model the forbidden line emission is dependent on accretion rate
and thus the model cannot explain the absence of strong late-type \OI
sources. The OH photodissociation model suggested in the present paper
naturally predicts that passive disks around weak UV sources will not
produce strong \OI 6300\A emission.

The high-velocity blue wings in the \OI 6300\A line of a small minority of
our sample stars cannot be accounted for by an emitting passive
Keplerian disk. This 
emission feature is suggested to emanate from an outflow, of which the
redshifted part is occulted by the circumstellar disk. Note however
that this pronounced blue wing is often accompanied by a symmetric peak at
low velocities (e.g., \object{Z CMa}, \object{PV Cep}, \object{V645
 Cyg}, \object{HD 200775}). The latter might again be formed in the
surface layers of the rotating disk. Alternatively, the model of
\citet{kwan} for T Tauri stars may be valid for these
objects: since the assumption of passivity is likely not to be
valid for the disks of the group~III members, these sources might resemble
classical accreting T Tauri stars. As noted before, accretion is needed
to create the right settings for the T Tauri model to work. This idea
is supported by the detection of the [\SII] 6731\A line in \object{Z
CMa}, which is exclusive to this target in the present sample. We
note however that this behaviour seems to be rather exceptional for
HAEBE stars.

The values for the fractional OH abundance $\epsilon(OH)$ needed to
explain the observed \OI 6300\A luminosities are $\sim$10$^{-7}-10^{-6}$.
Observations of diffuse interstellar clouds show that
relative OH abundances of this magnitude ($\sim$10$^{-7}$) occur in the
interstellar medium \citep[e.g.,][and references therein]{crutcher79}.
Nevertheless, these values are two orders larger than
the abundances computed in 
recent models including a full treatment of disk chemistry
\citep[e.g.,][]{markwick02,kamp04}. A possible reason for this 
discrepancy may be the exact location of the \OI emission region in our
models: a shift to higher densities (i.e. to a lower vertical height
$z$) would reduce the fractional OH abundance required to fit the
observed \OI intensity, since the latter is
inversely proportional to the first. This effect can potentially be
induced by the input dust opacities, which define the disk's flaring,
but also the UMIST coefficients determine the exact location of the
\OI emitting region through formula~(\ref{nunonth}). Alternatively,
the chemical models may be wrong due to uncertainties in the
reaction rate coefficients or the incompleteness of the
network.

\citet{bohm97} have stated that the forbidden-line emission region
in \textit{\citet{hillenbrand} group
I} sources\footnote{\citeauthor{hillenbrand} have defined their groups 
based on the near-IR spectral slope. There is no direct link between
both classifications: Hillenbrand group~I and II
contain Meeus group~I as well as group~II sources.} 
cannot be located at the 
surface of the circumstellar disk, because the suggested accretion
disk would cover most of the forbidden-line emission region.
The authors suggest that, even when the stellar light is able to reach
the disk surface, a problem remains: the upper layer would be
geometrically very thin and the outer radius of the disk
would need to be much larger than 100 AU in order to create enough
emission volume to explain the observed \OI intensities. Both problems
are countered when considering a flared disk with a puffed-up inner
rim. \citeauthor{hillenbrand} have invoked a circumstellar accretion disk
model to explain the observed near-IR ($\sim$2~$\mu$m) \textit{bump}
which is typical for Hillenbrand group~I sources. The puffed-up inner
rim in the \citet{dullemond01} model naturally
explains this excess. In other words, no dynamically active model is
needed to explain this bump. Furthermore, the outer parts of
the disk can be flared, hence increasing the angle $\alpha$ under
which the stellar light impinges onto the disk surface. The UV flux
can penetrate deeper into the disk, thus increasing the geometrical
thickness of the \OI emission layer. As we have shown in the present
paper, the observed \OI 6300\A emission profile can be explained as
being due to the photodissociation of OH molecules and the subsequent
non-thermal excitation of oxygen atoms in the atmosphere of a rotating
flared disk.

\section*{Appendix}

The fractional level population of two levels $a$ and $b$ of an atom in thermal
equilibrium at a temperature $T$[K] is given by
\begin{equation}
 \frac{N_{b}}{N_{a}} = \frac{g_b}{g_a} \frac{1}{1+\frac{A_{ba}}{n
    \gamma_{ba}}} \exp \left( -\frac{E_{ba}}{T} \right)  \label{thermalequi}
\end{equation}
In this formula, $g_a$ and $g_b$ are the statistical weights, $A_{ba}$
is the Einstein transition rate for the transition between level $b$ and $a$ and
$E_{ba}$ is the energy of the transition in K. The factor
$n$ is the number density of the collisional partner,
$\gamma_{ba}$ is the collisional rate for the transition and the
collisional partner. When more than one collisional partner is involved,
the quantity $n \gamma_{ba}$ is equal to the sum of these values for
each interacting species $i$: $n \gamma_{ba} = \Sigma_i\, n_i\,
\gamma_{ba}^i$. 

To determine the fraction $F$ of neutral oxygen in the $^1D_2$ state in the
five-level model assuming thermal excitation, one applies
\begin{equation}
 F = \left( \frac{\Sigma_{j=1}^5 N_j}{N_{^1D_2}} \right)^{-1} = \left(
 \Sigma_{j=1}^5 \frac{N_j}{N_{^1D_2}} \right)^{-1} 
\end{equation}
These ratios can easily be computed using equation
(\ref{thermalequi}), the atomic data and the fact that
\begin{equation}  
\frac{N_a}{N_b} = \left( \frac{N_b}{N_a} \right)^{-1} 
\end{equation}

\acknowledgements{
The authors would like to thank the support staff at La Silla and
Kitt Peak observatories for their excellent support during the
observing runs on which this paper is based. Especially the expertise
of G.~Lo~Curto (ESO La Silla) and D.~Willmarth (KPNO) proved
invaluable in completing our project succesfully. We thank the
anonymous referee for insightful comments which improved both contents
and presentation of the manuscript. BA would like to
thank I.~Kamp for the useful discussions concerning chemical
modeling.}

%------------------------------------------------------------------

\bibliographystyle{aa}
\bibliography{/STER/55/bram/REFERENCES/references.bib}

\end{document}